\documentclass[fleqn,usenatbib,usedcolumn]{mnras}

\usepackage[british]{babel}             
\usepackage{amsmath}
\usepackage{newtxtext}                  
   \usepackage[slantedGreek]{newtxmath}    
   \usepackage[T1]{fontenc}                
   \usepackage{graphicx}                   
\usepackage{pdflscape}
\usepackage{tabu}
\usepackage{xcolor}

\title[28 -- 40 GHz variability and polarimetry of bright compact sources in the QUIJOTE cosmological fields]{28 -- 40 GHz variability and polarimetry of bright compact sources in the QUIJOTE cosmological fields}

\author[Y. Perrott et al.]{
Yvette C. Perrott$^{1,2}$\thanks{E-mail: yvette.perrott@vuw.ac.nz}, Marcos L\'{o}pez-Caniego$^{3}$, \newauthor
Ricardo T. G\'{e}nova-Santos$^{4,5}$, Jose Alberto Rubi\~{n}o-Mart\'{i}n$^{4,5}$, Mark Ashdown$^{1,6}$, \newauthor
Diego Herranz$^{7}$, Anne L\"ahteenm\"aki$^{8,9}$, Anthony N. Lasenby$^{1,6}$, \newauthor
Carlos H. L\'{o}pez-Caraballo$^{4,5,10}$, Fr\'{e}d\'{e}rick Poidevin$^{4,5}$, Merja Tornikoski$^{8}$
\\
$^{1}$Astrophysics Group, Cavendish Laboratory, JJ Thomson Avenue, Cambridge CB3 0HE, UK\\
$^{2}$School of Chemical and Physical Sciences, Victoria University of Wellington, PO Box 600, Wellington 6140, New Zealand\\
$^{3}$European Space Agency, ESAC, Camino bajo del Castillo, s/n, Urbanizaci\'{o}n Villafranca del Castillo, Villanueva de la Ca\~{n}ada, E-28692 Madrid, Spain\\
$^{4}$Instituto de Astrof\'{i}sica de Canarias (IAC), E-38205 La Laguna, Tenerife, Spain\\
$^{5}$Universidad de La Laguna, Dpto. Astrof\'{i}sica, E-38206 La Laguna, Tenerife, Spain\\
$^{6}$Kavli Institute for Cosmology, University of Cambridge, Madingley Road, Cambridge CB3 0HA, UK\\
$^{7}$Instituto de F\'{i}sica de Cantabria (CSIC-Universidad de Cantabria), Avda. de los Castros s/n, E-39005 Santander, Spain\\
$^{8}$Aalto University Mets\"ahovi Radio Observatory, Mets\"ahovintie 114, 02540 Kylm\"al\"a, Finland\\
$^{9}$Aalto University Department of Electronics and Nanoengineering, P.O. BOX 15500, FI-00076 AALTO, Finland\\
$^{10}$INFN Sezione di Milano, Via Celoria 16, Milano, Italy\\
}

\date{Accepted 2021 February 02. Received 2021 February 01; in original form 2020 February 03}

\pubyear{2021}

\makeatletter
\def\convertto#1#2{\strip@pt\dimexpr #2*65536/\number\dimexpr 1#1}
\makeatother

\begin{document}
\label{firstpage}
\pagerange{\pageref{firstpage}--\pageref{lastpage}}
\maketitle

\begin{abstract}

We observed 51 sources in the Q-U-I JOint TEnerife (QUIJOTE) cosmological fields which were brighter than 1\,Jy at 30\,GHz in the \emph{Planck} Point Source Catalogue (version 1), with the Very Large Array at 28 -- 40\,GHz, in order to characterise their high-radio-frequency variability and polarization properties.  We find a roughly log-normal distribution of polarization fractions with a median of 2\%, in agreement with previous studies, and a median rotation measure (RM) of $\approx$\,1110\,rad\,m$^{-2}$ with one outlier up to $\approx$\,64000\,rad\,m$^{-2}$ which is among the highest RMs measured in quasar cores.  We find hints of a correlation between the total intensity flux density and median polarization fraction.  We find 59\% of sources are variable in total intensity, and 100\% in polarization at $3\sigma$ level, with no apparent correlation between total intensity variability and polarization variability.  This indicates that it will be difficult to model these sources without simultaneous polarimetric monitoring observations and they will need to be masked for cosmological analysis.

\end{abstract}



\section{Introduction}

The Q-U-I Joint TEnerife (QUIJOTE) experiment \citep{2010ASSP...14..127R} aims to detect inflationary B-modes using CMB observations at 31 and 42\,GHz with supporting observations between 10 -- 20\,GHz to characterise foregrounds.  The low angular resolution ($\approx$\,$1^{\circ}$) of the instrument means that polarized extragalactic radio sources, a major contaminant at small angular scales (e.g.\ \citealt{2018ApJ...858...85P}), cannot be accurately characterised by the experiment and ancillary data must be used to either subtract or mask them.  Ideally these sources would be monitored at higher resolution simultaneously with the cosmological observations as for the Very Small Array experiment \citep{2003MNRAS.341L..23G}, however since this was not possible we investigated the long-term properties of the bright sources in the field to assess the level of contamination an inaccurate subtraction would introduce.  The polarization and variability properties of radio sources at these frequencies are not well-studied; some of the relevant existing studies are listed in Table~\ref{T:other_studies}.  Polarimetric very-long-baseline interferometric monitoring studies exist at 43\,GHz, e.g.\ \citet{2018ApJ...860..112P}, \citet{2011MNRAS.411...85A}, \citet{2012MNRAS.420..542A}, \citet{2007AJ....134..799J}, \citet{2001ApJ...562..208L}; however these focus on spatially resolving properties of the sources rather than investigating their integrated properties, which are of relevance to experiments such as QUIJOTE.  At the other end of the resolution scale, \citet{2017MNRAS.469.2401B} stack \emph{Planck} data to statistically detect polarization from known source positions and constrain average polarization properties, but are too limited by sensitivity to investigate variability.  \citet{2016IJMPD..2540005G}, \citet{2017MNRAS.465.4085G} and \citet{2018MNRAS.475.1306G} detect a large sample of sources between 2 -- 38\,GHz, however they only have repeated observations for a small subsample of sources and focus on broadband polarimetry rather than variability.  It is clear that there are very few studies addressing the integrated variability of bright, polarized sources at 30 -- 40\,GHz; we therefore observed the brightest sources in the QUIJOTE fields using the Very Large Array (VLA) to determine at what level in intensity a source could be safely subtracted in polarization.

\begin{landscape}
\begin{table}
\centering
\caption{Characteristics of polarimetric studies of radio sources at frequencies overlapping the 30 -- 40\,GHz band of this study.}
\label{T:other_studies}
\begin{tabu} to \textwidth {XcXXXX[2]}
\hline
Reference & Instrument & Frequencies & Epochs & Number of sources & Comment\\ \hline
\citet{1985ApJS...57..693R}, \citet{1985ApJ...290..627J} & 1.4 -- 90\,GHz (8 bands) & Kitt Peak 10.6\,m (for 31\,GHz) & 6 total, 2 at 31\,GHz & $\approx$\,20 & Strong, flat-spectrum sources \\[5pt]
\citet{2009ApJ...705..868L} & WMAP & 23, 33, 41, 61\,GHz & Single 5-year average & 41 (30) at 99\% significance at 33 (41)\,GHz & Survey maps correlated with total intensity source catalogue \\[5pt]
\citet{2010MNRAS.401.1388J}, \citet{2011MNRAS.413..132B} & VLA & 8, 22, 43\,GHz & 1 & 203 & Brighter than 1\,Jy in the WMAP 22\,GHz catalogue \\[5pt]
\citet{2011ApJ...732...45S} & VLA & 4, 8, 22, 43\,GHz & 1 at 4, 8, 43\,GHz; 2 at 22\,GHz & 159 & Radio galaxies with Australia Telescope 20 GHz (AT20G) survey flux densities $>40$\,mJy in Atacama Cosmology Telescope cosmological field \\[5pt]
\citet{2015IAUS..313..128K} & VLA & 1.4 -- 43\,GHz (8 bands) & 3 & 7 & Bright calibration sources \\[5pt]
\citet{2015ApJ...806..112H} & Q/U Imaging Experiment & 43, 95\,GHz & Single 25-month average & 13 at S/N$>2.7$ at 43\,GHz & Survey maps correlated with AT20G sources \\[5pt]
\citet{2016IJMPD..2540005G}, \citet{2017MNRAS.465.4085G}, \citet{2018MNRAS.475.1306G} & ATCA & 2 -- 38\,GHz & 2 in polarization for 10 sources & 104 & `Faint' Planck-ATCA Coeval Observation sample, $>200$\,mJy at 20\,GHz, Southern Ecliptic Pole region \\[5pt]
\citet{2017MNRAS.469.2401B}, \citet{2016AA...594A..26P} & \emph{Planck} & 30 -- 353\,GHz & 4 years, averaged & 1560 (stacked) & PCCS2 30\,GHz sources $>427$\,mJy \\
\hline
\end{tabu}
\end{table}
\end{landscape}

The paper is organised as follows.  In Section~\ref{S:sample} we introduce our sample; in Section~\ref{S:obs} we describe our observations, data reduction and analysis methods.  In Section~\ref{S:results} we summarize our results and investigate correlations between the measured quantities.  In Section~\ref{S:ind_sources} we investigate in more detail some interesting or anomalous cases.  In Section~\ref{S:implications} we discuss implications for QUIJOTE and in Section~\ref{S:conclusions} we conclude.

Throughout this paper we will use the following conventions.  When fitting a power-law as a function of frequency we will use the convention $S \propto \nu^{\alpha}$ (i.e.\ $\alpha>0$ means a rising function of frequency).  For polarization, we will use $\mathcal{I}$, $\mathcal{Q}$, $\mathcal{U}$ to denote flux density in total intensity and Stokes Q and U respectively.  We will use $q = \mathcal{Q}/\mathcal{I}$ and $u = \mathcal{U}/\mathcal{I}$ to indicate the fractional quantities and similarly $\mathcal{P} = \sqrt{\mathcal{Q}^2+\mathcal{U}^2}$ and $p = \mathcal{P}/\mathcal{I}$ to denote the total and fractional linear polarization.  Polarization angle will be indicated as $\Phi$, i.e.\ $\tan(2\Phi) = u/q$.  Circular polarization will be assumed to be zero throughout.

\section{Sample properties}
\label{S:sample}

We selected the 54 sources with flux densities greater than 1\,Jy at 30\,GHz in the first \emph{Planck} Catalogue of Compact Sources (PCCS1, \citealt{2014A&A...571A..28P}) lying within 30$^{\circ}$ of the centres of the three QUIJOTE cosmological fields at RA, $\delta$ = (00h40m, +25$^{\circ}$), (09h40m, +45$^{\circ}$) and (16h20m, +50$^{\circ}$) respectively.  We note that the second, more-complete version of the catalogue (PCCS2, \citealt{2016AA...594A..26P}) was not available at the time of sample selection however completeness is not an issue for this high-signal-to-noise-ratio (SNR) sample since at 30\,GHz PCCS1 is estimated to be 90\% complete at 575\,mJy.  Had we selected from PCCS2, $\approx$\,10\% of the sample would have been different due to the source flux densities varying over/under the 1\,Jy threshold.

Three of the sources in the sample were excluded due to large angular size, leaving a sample of 51 which were observed with the VLA at frequencies between 28 and 40\,GHz, with 41 sources having two epochs of observation.  The sources, as expected given the high-frequency selection, are mostly blazars and have identifications in the 5th edition of the Roma-BZCAT Multifrequency Catalogue of Blazars \citep{2015Ap&SS.357...75M}; these are listed in Table~\ref{T:sample} along with coordinates and other identifications taken from the SIMBAD Astronomical Database\footnote{\url{http://simbad.u-strasbg.fr/simbad/}}.  The two sources not in the BZCAT are a LINER-type AGN and a Seyfert-1 galaxy.

\begin{table*}
\begin{center}
\caption{Sources in the sample, positions and associations taken from SIMBAD, classifications taken from BZCAT \citep{2015Ap&SS.357...75M}.  Redshifts are from BZCAT and references therein unless otherwise specified.  The three resolved sources are excluded from the VLA sample.}
\label{T:sample}
\scriptsize
\begin{tabu} to \linewidth{lcccccX}
\hline
PCCS1\,030 & $z$ & RA & Dec & Name & Class & Comment\\
\hline
G107.00-50.62 & 0.089 & 00:10:31.01 & +10:58:29.5 & 5BZQ J0010+1058 & FSRQ & Mrk 1501 (Sy1)\\
G115.13-64.85 & 0.21968$^a$ & 00:38:20.53 & -02:07:40.5 &  &  & 3C 17 (resolved) \\
G124.55-32.50 & 0.016485$^b$ & 00:57:48.89 & +30:21:08.8 &  &  & NGC 315 (LIN)\\
G131.84-60.98 & 2.099 & 01:08:38.77 & +01:35:00.3 & 5BZQ J0108+0135 & FSRQ & 4C 01.02, [V2003b] QSO J0108+0135 (QSO with absorption lines)\\
G129.46-49.29 & 0.05970$^c$ & 01:08:52.87 & +13:20:14.3 &  &  & 3C 33 (resolved) \\
G134.59-50.34 & 0.57 & 01:21:41.60 & +11:49:50.4 & 5BZQ J0121+1149 & FSRQ & QSO B0119+115\\
G129.09-13.46 & 0.067 & 01:28:08.06 & +49:01:06.0 & 5BZG J0128+4901 & BL Lac-galaxy dominated & 2MASX J01280804+4901056 (Sy1)\\
G130.78-14.31 & 0.859 & 01:36:58.59 & +47:51:29.1 & 5BZQ J0136+4751 & FSRQ & QSO B0133+476\\
G141.09-38.56 & 1.32 & 01:52:18.06 & +22:07:07.7 & 5BZQ J0152+2207 & FSRQ & QSO B0149+218\\
G147.84-44.04 & 0.833 & 02:04:50.41 & +15:14:11.0 & 5BZU J0204+1514 & Blazar Uncertain type & 4C 15.05 (blazar)\\
G140.56-28.12 & 1.466 & 02:05:04.93 & +32:12:30.1 & 5BZQ J0205+3212 & FSRQ & 7C 020209.69+315812.00 (QSO)\\
G152.57-47.29 & 0.20$^d$ & 02:11:13.18 & +10:51:34.8 & 5BZB J0211+1051 & BL Lac & 2MASS J02111317+1051346 (BLL)\\
G140.23-16.73 & 0.021258$^e$ & 02:23:11.41 & +42:59:31.4 &  &  & UGC 1841 (Sy1)\\
G157.08-42.72 & 2.065 & 02:31:45.89 & +13:22:54.7 & 5BZQ J0231+1322 & FSRQ & 4C 13.14, [RKV2003] QSO J0231+1322 (QSO with absorption lines)\\
G149.50-28.53 & 1.206 & 02:37:52.41 & +28:48:09.0 & 5BZQ J0237+2848 & FSRQ & 4C 28.07 (QSO)\\
G156.86-39.13 & 0.94 & 02:38:38.93 & +16:36:59.3 & 5BZB J0238+1636 & BL Lac & [RKV2003] QSO J0238+1637, QSO B0235+1624, [CGK2006] QSO 0235+164 -00.33-0.45 (QSO with ALS, BLL, G?)\\
G178.26+33.40 & 0.53 & 08:18:16.00 & +42:22:45.4 & 5BZB J0818+4222 & BL Lac & 7C 081451.89+423206.00 (BLL)\\
G182.17+34.17 & 1.216 & 08:24:55.48 & +39:16:41.9 & 5BZQ J0824+3916 & FSRQ & 4C 39.23A (QSO)\\
G200.04+31.88 & 0.939 & 08:30:52.09 & +24:10:59.8 & 5BZQ J0830+2410 & FSRQ & [RKV2003] QSO J0830+2411, 7C 082754.29+242103.00 (QSO with absorption lines)\\
G143.53+34.42 & 2.218 & 08:41:24.36 & +70:53:42.2 & 5BZQ J0841+7053 & FSRQ & 7C 083620.60+710405.00, [TOS2004] QSO J0841+7053 (QSO with absorption lines)\\
G206.82+35.81 & 0.306 & 08:54:48.87 & +20:06:30.6 & 5BZB J0854+2006 & BL Lac & QSO J0854+2006 (BLL)\\
G175.72+44.81 & 2.19 & 09:20:58.46 & +44:41:54.0 & 5BZQ J0920+4441 & FSRQ & 7C 091741.89+445438.00 (QSO)\\
G152.23+41.00 & 1.446 & 09:21:36.23 & +62:15:52.2 & 5BZQ J0921+6215 & FSRQ & 7C 091740.39+622838.00 (QSO)\\
G198.82+44.43 & 0.744 & 09:23:51.52 & +28:15:25.0 & 5BZQ J0923+2815 & FSRQ & 9C J0923+2815 (QSO)\\
G183.71+46.17 & 0.695 & 09:27:03.01 & +39:02:20.9 & 5BZQ J0927+3902 & FSRQ & ICRF J092703.0+390220 (Sy1)\\
G181.02+50.29 & 1.249 & 09:48:55.34 & +40:39:44.6 & 5BZQ J0948+4039 & FSRQ & 7C 094550.20+405345.00 (QSO)\\
G141.43+40.58 & 0.000677$^f$ & 09:55:52.73 & +69:40:45.8 &  &  & M82 (resolved)   \\
G145.78+43.13 & 0.367 & 09:58:47.25 & +65:33:54.8 & 5BZB J0958+6533 & BL Lac & 7C 095457.89+654812.00 (BLL)\\
G177.37+58.35 & 1.117 & 10:33:03.71 & +41:16:06.2 & 5BZQ J1033+4116 & FSRQ & 7C 103008.00+413135.00 (QSO)\\
G211.56+60.99 & 0.56 & 10:43:09.03 & +24:08:35.4 & 5BZQ J1043+2408 & FSRQ & 2MASS J10430901+2408354 (BLL)\\
G135.47+42.26 & 1.15 & 10:48:27.62 & +71:43:35.9 & 5BZU J1048+7143 & Blazar Uncertain type & 7C 104450.00+715929.00 (QSO)\\
G135.91+43.92 & 2.492 & 10:56:53.62 & +70:11:45.9 & 5BZQ J1056+7011 & FSRQ & 7C 105328.60+702751.00 (QSO)\\
G133.79+42.34 & 1.462 & 11:01:48.81 & +72:25:37.1 & 5BZQ J1101+7225 & FSRQ & 7C 105819.89+724145.00 (QSO)\\
G174.43+69.81 & 1.733 & 11:30:53.28 & +38:15:18.5 & 5BZQ J1130+3815 & FSRQ & QSO B1128+385 (QSO)\\
G145.65+64.96 & 0.334 & 11:53:24.47 & +49:31:08.8 & 5BZQ J1153+4931 & FSRQ & 7C 115048.00+494751.00 (QSO)\\
G098.27+58.31 & 0.153 & 14:19:46.60 & +54:23:14.8 & 5BZB J1419+5423 & BL Lac & 7C 141805.29+543657.00 (BLL)\\
G105.20+49.72 & 2.068 & 14:36:45.80 & +63:36:37.9 & 5BZQ J1436+6336 & FSRQ & [RKV2003] QSO J1436+6336, 8C 1435+638, [TOS2004] QSO J1436+6336 (QSO with absorption lines)\\
G055.14+46.37 & 1.397 & 16:13:41.06 & +34:12:47.9 & 5BZQ J1613+3412 & FSRQ & 7C 161147.89+342019.00 (QSO)\\
G061.07+42.34 & 1.814 & 16:35:15.49 & +38:08:04.5 & 5BZQ J1635+3808 & FSRQ & [V2003b] QSO J1635+3808, 7C 163330.69+381410.00 (QSO with absorption lines)\\
G073.40+41.88 & 0.735 & 16:37:45.13 & +47:17:33.8 & 5BZQ J1637+4717 & FSRQ & 7C 163619.69+472344.00 (QSO)\\
G086.64+40.35 & 0.751 & 16:38:13.46 & +57:20:24.0 & 5BZQ J1638+5720 & FSRQ & ICRF J163813.4+572023 (Sy1)\\
G100.68+36.62 & 0.751 & 16:42:07.85 & +68:56:39.7 & 5BZQ J1642+6856 & FSRQ & 6C 164218+690201 (QSO)\\
G063.46+40.96 & 0.593 & 16:42:58.81 & +39:48:37.0 & 5BZQ J1642+3948 & FSRQ & 3C 345, 7C 164117.60+395412.00 (QSO)\\
G071.46+33.28 & 0.717 & 17:27:27.65 & +45:30:39.7 & 5BZQ J1727+4530 & FSRQ & ICRF J172727.6+453039 (Sy1)\\
G064.03+31.01 & 0.976 & 17:34:20.58 & +38:57:51.4 & 5BZQ J1734+3857 & FSRQ & 7C 173240.70+385949.00 (BLL)\\
G079.56+31.72 & 1.381 & 17:40:36.98 & +52:11:43.4 & 5BZQ J1740+5211 & FSRQ & 7C 173928.89+521311.00 (QSO)\\
G100.13+29.16 & 0.046 & 18:06:50.68 & +69:49:28.1 & 5BZB J1806+6949 & BL Lac & 7C 180717.89+694858.00 (BLL)\\
G085.73+26.11 & 0.663 & 18:24:07.07 & +56:51:01.5 & 5BZB J1824+5651 & BL Lac & 7C 182315.19+564917.00 (BLL)\\
G077.23+23.50 & 0.695 & 18:29:31.80 & +48:44:46.7 & 5BZU J1829+4844 & Blazar Uncertain type & 2C 1569 (Sy1)\\
G097.50+25.03 & 0.657 & 18:49:16.08 & +67:05:41.7 & 5BZQ J1849+6705 & FSRQ & ICRF J184916.0+670541 (Sy1)\\
G090.09-25.64 & 0.79 & 22:36:22.47 & +28:28:57.4 & 5BZQ J2236+2828 & FSRQ & QSO B2232+282 (BLL)\\
G086.10-38.18 & 0.859 & 22:53:57.75 & +16:08:53.6 & 5BZQ J2253+1608 & FSRQ & 3C 454.3, [RKV2003] QSO J2253+1608 (QSO with absorption lines)\\
G108.97-09.47 & 1.279 & 23:22:26.00 & +50:57:52.0 & 5BZQ J2322+5057 & FSRQ & ICRF J232225.9+505751 (QSO)\\
G091.12-47.97 & 1.843 & 23:27:33.58 & +09:40:09.5 & 5BZQ J2327+0940 & FSRQ & QSO B2325+093\\
\hline
\end{tabu}
\end{center}
\noindent$^a$\citet{1965ApJ...141....1S}, $^b$\citet{2000AJ....119.1645T}, $^c$\citet{1996A&AS..116...43P}, $^d$\citet{2010ApJ...712...14M}, $^e$\citet{1999ApJS..121..287H}, $^f$\citet{1991rc3..book.....D}
\end{table*}

\section{Observations and data reduction}\label{S:obs}

The sources were observed on the dates listed in Table~\ref{T:obs_dates}.  Observations were carried out with a custom correlator configuration which allowed simultaneous observing at two frequency bands, 28.5 -- 32.4 and 35.5 -- 39.4 GHz.  Each band was then divided into 32 spectral windows, each with 64 channels of 2\,MHz width.

\begin{table}
\begin{center}
\caption{Dates on which the sources were observed.  `Num.' refers to the number of target sources observed in each observation.  `Beam size' is the average synthesized beam major and minor full width at half maximum, in arcsec.}\label{T:obs_dates}
\begin{tabu} to \linewidth{lccccc}
\hline
Project & Date & Num. & Flux & VLA & Beam \\
code & & & calibrator & config. & size \\ \hline
15A-083 & 10/03/2015 & 19 & 3C\,48 & B & $0.3 \times 0.2$ \\
15A-083 & 15/03/2015 & 17 & 3C\,48 & B & $0.4 \times 0.2$ \\
16A-035 & 15/02/2016 & 18 & 3C\,286 & C & $1.2 \times 0.7$ \\
16A-035 & 11/03/2016 & 11 & 3C\,48 & C & $1.3 \times 0.7$ \\
17A-027 & 16/06/2017 & 8 & 3C\,48 & C & $0.8 \times 0.7$ \\ 
18B-003 & 18/11/2018 & 8 & 3C\,286 & C & $0.9 \times 0.7$ \\
18B-003 & 27/11/2018$^{a}$ & 5 & 3C\,286 & C & $0.9 \times 0.7$ \\
18B-003 & 14/12/2018 & 4 & 3C\,286 & C & $1.0 \times 0.7$ \\
18B-003 & 16/12/2018$^{b}$ & 5 & 3C\,286 & C & $0.8 \times 0.7$ \\
18B-003 & 28/01/2019 & 6 & 3C\,286 & C & $1.4 \times 0.7$ \\\hline
\multicolumn{6}{l}{$^{a}$ Data affected by instrumental issues and not used} \\
\multicolumn{6}{l}{$^{b}$ Repeat of 27/11/2018 observation} \\
\end{tabu}
\end{center}
\end{table}

We used either 3C\,286 or 3C\,48 as the flux density calibrator, both in total intensity and polarization, depending on which was closer in the sky to the sources being observed.  In the case of 3C\,286, we used J1407+2827 as the polarization leakage calibrator; in the case of 3C\,48 we used 3C\,84.  These are both recommended as primary low polarization leakage calibrators for the VLA (\url{https://science.nrao.edu/facilities/vla/docs/manuals/obsguide/modes/pol}; see Table 7.2.3).  Since these sources are blazars, their emission at these frequencies is dominated by the compact core and they can be treated to a good approximation as point sources.  We therefore used the targets themselves as phase calibrators, and performed pointing calibration scans at X-band throughout the observations on targets less than 20$^{\circ}$ from the subsequent scans.

The data reduction was performed in \textsc{casa} 5.3.0 and was based on the standard VLA data reduction pipeline adapted for our observations.  As in the pipeline, we use the standard flux calibration sources 3C\,48 or 3C\,286 to calculate delay and bandpass calibrations.  Departing from the pipeline, we then assume a point source model for each of the target sources and self-calibrate them in amplitude and phase.  We use the `fluxscale' task to translate the amplitude solutions into a spectral fit for each target source; see Fig.~\ref{Fi:sample_pol_fits} for an example fit).

To check that our point source model accurately describes the sources, we subtract the model from the calibrated data and inspect the residual maps and visibilities.  In most cases the residuals are $<1$\% of the core flux density.  In the few cases that the residuals are larger, we iteratively improve the model by adding in the non-core flux observed on the map, then solving again for the core flux density in the updated core $+$ non-core model.  Maps and $uv$-plane residuals are shown for all sources in Appendix~A (available as supplementary material).

We then follow EVLA Memo 201 \citep{EVLAM201} and \citet{2017AJ....154...54H} for the polarization calibration strategy, in particular following strategy `C1' with one unpolarized calibrator and one calibrator with known polarization:

\begin{itemize}
\item{Set the full polarization model for the flux density calibrator using fits to polarization fractions and angle as a function of frequency from \citet{2013ApJS..206...16P}.  If using 3C\,48, which is resolved, flag baselines where the model \emph{I} flux is less than 90\% of the total model \emph{I} flux and treat as an unresolved source}
\item{Find cross-hand delay (`\textsc{kcross}') solutions using calibrator with known polarization}
\item{Solve for polarization leakage per-channel (`\textsc{Df}') using the unpolarized calibrator if bright enough; otherwise (observations using J1407+2827 only) solve using a combination of the science targets (see below for more details)}
\item{Solve for the absolute polarization angle (`\textsc{Xf}') using calibrator with known polarization.}
\end{itemize}

As for the total intensity measurements, we solve for the spectral energy distributions of the sources in polarization in the \emph{uv}-plane; this will be described in Section~\ref{S:qu_fitting}.  We also check the image and $uv$-plane for significant residuals from the point-source model and find none with the exception of the source PCCS1~G182.17+34.17, which displays significant polarization residuals at the positions of the intensity peaks seen in Fig.~A42.  For consistency with the other sources we quote the point-source fit for this source, however clearly when observed by a lower-resolution instrument such as QUIJOTE this source would appear to have a different polarization.

In the 2018 -- 2019 observations, the polarization leakage calibrator J1407+2827 had become too faint to successfully solve for the per-channel leakage solutions.  We therefore adapted our calibration procedure as follows.  We made a first-pass leakage calibration with J1407+2827 using per-spectral-window (`\textsc{D}') solutions.  We then used all science targets with polarization fractions $<5\%$ (from the first-pass analysis), plus 3C286, to make a second-pass solution, as follows.  We used the first-pass solution to insert a model for the polarized flux for all sources.  We then repeated the polarization leakage calibration using `\textsc{Df+X}' mode to solve simultaneously for the leakage and polarization angle terms.  We expect any errors in the first-pass solution to leave only a small polarized residual which will tend to cancel out between the different sources since the polarization angles are uncorrelated.  We then solve for the polarization angle using 3C286 as usual.

\begin{figure}
\begin{tabular}{c}
\includegraphics[width=\linewidth, keepaspectratio]{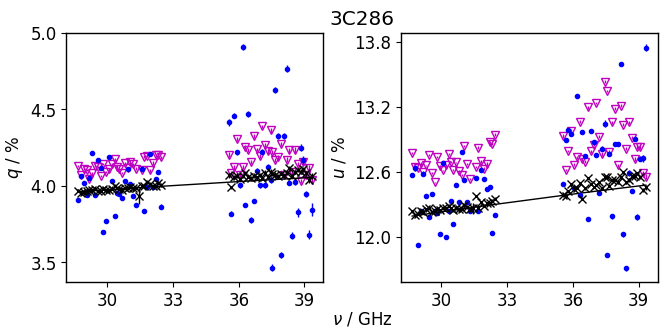}\\
\includegraphics[width=\linewidth, keepaspectratio]{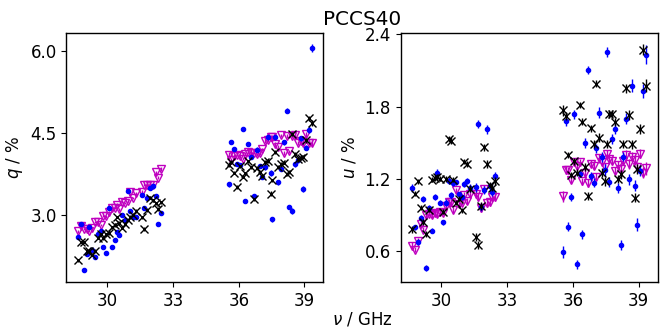}\\
\end{tabular}
\caption{$q$ and $u$ values recovered in each spectral window for two sources in the 14/12/2018 observation, after full calibration using (i) `\textsc{D}' solutions using J1407+2827 (blue dots); (ii) `\textsc{Df+X}' solutions using all science targets with $p_{34}<5$\% (black crosses); (iii) `\textsc{Df}' solutions using PCCS1\,G174.43+69.81 with a low $p_{34}=0.4$\% (magenta triangles).  In the 3C286 plot the black line is the model.  The `\textsc{Df+X}' solutions clearly lead to a more accurate calibration which is unbiased compared to the model and the (less accurate) J1407+2827 calibration; the PCCS1\,G174.43+69.81 calibration is biased even though the polarization fraction is very small.}\label{Fi:18B_polcal}
\end{figure}

To demonstrate the validity of this method, in Fig.~\ref{Fi:18B_polcal} we show a comparison between some sources calibrated using this method and with the `\textsc{D}' solutions using J1407+2827; it is clear that the results are the same but the calibration is more accurate.  We also attempted a more standard calibration by merely substituting J1407+2827 with the source with the lowest polarization fraction, as determined using the J1407+2827 `\textsc{D}' solutions.  We show an example calibration of this type in Fig.~\ref{Fi:18B_polcal} as well; although the new leakage calibration source has a very low polarization fraction of $p_{34}=0.4$\% it is still detected and using it as the leakage calibrator gives results that are biased with respect to the J1407+2827 and combined solutions.  Our `combined' method is therefore the best way to calibrate the polarization leakage given the limitations of the data.  We apply this method also to the 2016 observation using J1407+2827 as the leakage calibrator to improve the SNR of the leakage solutions.

\clearpage

\subsection{Calibration accuracy}\label{S:calibration_accuracy}

The expected calibration accuracy is 5\% given that we are using the 3-bit samplers and therefore cannot apply the switched power calibration\footnote{See \url{https://science.nrao.edu/facilities/vla/docs/manuals/oss/performance/vla-samplers\#section-1} for an explanation of the expected calibration accuracy.}.  Given the variable nature of the sources, assessment of the calibration accuracy requires simultaneous monitoring.  We use two datasets for this purpose; the Mets\"{a}hovi 37\,GHz blazar monitoring data \citep{1992A&AS...94..121T} and the Owens Valley Radio Observatory (OVRO) blazar monitoring data at 15\,GHz \citep{2011ApJS..194...29R}; since many of our sources are frequently monitored by both programmes we can interpolate the monitoring flux densities to the appropriate time and compare to our VLA measurements.

\begin{figure}
\centering
\includegraphics[width=0.65\linewidth, keepaspectratio]{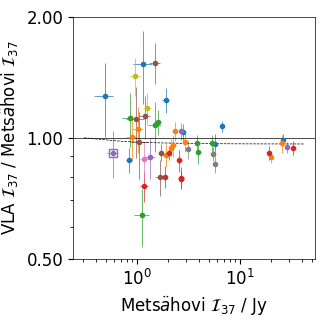}
\includegraphics[width=0.65\linewidth, keepaspectratio]{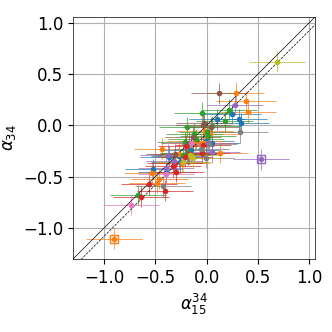}
\caption{Simultaneous 37\,GHz total intensity flux densities measured by Mets\"{a}hovi and the VLA (top), and simultaneous 15 -- 34\,GHz spectral indices, $\alpha_{15}^{34}$ calculated using the OVRO blazar monitoring data, compared to spectral indices across the VLA bands, $\alpha_{34}$ (bottom).  The points are coloured by VLA observing epoch.  In both cases, the solid black line shows a one-to-one correspondence while the black dashed line shows the best straight-line fit using orthogonal least-squares regression.  In the Mets\"{a}hovi plot, no extra systematic errors have been added to the VLA datapoints.  In the OVRO plot, conservative VLA calibration uncertainties of 5\% in the reference flux density and 0.1 in the spectral index are included in the errorbars plotted and used for the least-squares fit.  The points marked with squares in both plots are PCCS1\,G107.00-50.62 which may have a turnover in its spectrum considering the 37\,GHz flux agrees well but the spectral index is discrepant with OVRO.  These points are excluded from the fit to the OVRO data.}\label{Fi:Mets_OVRO}
\end{figure}

In the case of Mets\"{a}hovi, we find 48 semi-simultaneous (within 10 days) measurements of 30 sources, including the calibration source 3C\,84.  We correct our VLA measurements to 37\,GHz using the fitted in-band spectral index and compare the measured flux densities; this is shown in Fig.~\ref{Fi:Mets_OVRO}.  In general we see good agreement.  There is a slight overabundance of sources with significantly low VLA flux density compared to Mets\"{a}hovi which probably indicates a resolution effect.  The Mets\"{a}hovi beam is 2.4\,arcmin at 37\,GHz.  The VLA primary beam across our band has a FWHM of at most 1.6\,arcmin, and the shortest VLA baselines are $\approx$\,10\,k$\lambda$ measuring structure on $\sim$\,20\,arcsec scales, so both nearby sources outside of the VLA primary beam and flux resolved on scales between 20\,arcsec and 2.4\,arcmin would be included in the Mets\"{a}hovi measurements but not detectable on our VLA maps.  In addition, the Mets\"{a}hovi errorbars are relatively large (3\% systematic error for the brightest sources, and errors up to 20\% for the faintest in the overlapping sample), and the lightcurves are often sparsely sampled (around 20 day average cadence). The combination of resolution, large errorbars and sparse sampling makes testing for the expected $\sim$\,5\% VLA calibration uncertainty difficult using the Mets\"{a}hovi data.  Fitting a straight line to $S_{\mathrm{VLA}}$ vs $S_{\mathrm{Mets\"{a}hovi}}$ using orthogonal least-squares regression gives a slope and offset consistent with $1$ and $0$ within $2\sigma$, where $\sigma$ are the errorbars of the fit.  We also note that the measurements of the brightest source, 3C\,84, are consistent within $2\sigma$ (without adding additional calibration errors to the VLA data) and conclude that the VLA and Mets\"{a}hovi data are consistent within the limitations of the data.

In the case of OVRO, we find 83 semi-simultaneous measurements of 43 sources.  To compare the VLA and OVRO flux measurements, we calculate spectral indices between 15 and 34\,GHz.  Assuming that there is no spectral curvature between 15 and 40\,GHz, this should be equal to the spectral index over the VLA bands.  Fig.~\ref{Fi:Mets_OVRO} shows that there is indeed a very good correspondence between these quantities.  We assume offsets from this relationship will be dominated by the VLA systematic calibration uncertainties in $\alpha_{34}$ due to the much shorter frequency lever arm and therefore use this comparison to test for the value of this uncertainty.  We do this by fitting a straight line to the data (excluding the significant outlier PCCS1\,G107.00-50.62) using orthogonal least-squares regression.  We use the OVRO errors as given in the database and use error propagation to calculate the error in the interpolated quantity.  We add an extra VLA calibration error in $\alpha_{34}$ to the statistical error calculated from the fit over the VLA band, and increase this error until reduced $\chi^2=1$ for the linear fit. This gives a spectral index calibration error of 0.1, which will be a conservative estimate given that we have not included any VLA systematic flux density calibration errors.  The orthogonal least-squares regression, with the nominal 5\% flux density calibration uncertainty and spectral index calibration uncertainty of 0.1 added to the statistical errors in quadrature, gives a best-fit slope consistent with one ($1.00 \pm 0.06$) and some indication of an offset ($-0.08 \pm 0.02$).  This may indicate a small systematic offset in the spectral indices fit across the VLA band; the effect of some sources having flux resolved out as described above for the Mets\"{a}hovi comparison; or a true slight steepening on average of the spectral indices from 15\,GHz to 40\,GHz.  We note that \citet{2011ApJ...732...45S} found a steepening of spectral indices between 20 and 90\,GHz and \citet{2016A&A...596A.106P} found broken power-law fits were necessary between 1.1 -- 857\,GHz with the high-frequency spectral index typically steeper than the low-frequency spectral index.  Given these uncertainties we choose not to attempt to correct for this offset.  We adopt the 5\% and 0.1 as the VLA systematic errors in central flux density and spectral index in the following variability analysis and add them in quadrature to the errors estimated from the fit to the VLA spectra; these are conservative upper limits to the true calibration accuracy as shown by the clearly overestimated total errorbars in Fig.~\ref{Fi:Mets_OVRO}.  We note that the two notable outliers (marked with squares in Fig.~\ref{Fi:Mets_OVRO}) are two measurements of the same source, PCCS1\,G107.00-50.62.  The single coincident VLA and 37\,GHz Mets\"{a}hovi measurement of this source agrees well, and it shows no non-core flux in the VLA maps (see Fig.~A1) so it may instead have a turnover in its spectrum accounting for the discrepancy.

To estimate the systematic uncertainties in the polarized quantities, we follow \citet{2017AJ....154...54H}.  For our `unpolarized' calibrators used for leakage calibration, we estimate the uncertainty in the polarization fraction $p_{\epsilon}$ as the spurious on-axis fractional polarization produced by a small amount of polarization actually being present:

\begin{align}
2 \left ( \sigma_{d} \right )^{2} &\approx p_{\mathrm{true}}^{2} + \frac{N_{a}}{A^2}\nonumber \\
p_{\epsilon} &\approx \sigma_{d} \sqrt{\frac{\pi}{2N_a}}
\end{align}
(using equations 30 and 28 from \citealt{2017AJ....154...54H}), where $\sigma_{d}$ is the error in the leakage `d-term' calibration accuracy, $N_a$ is the number of antennas, and $A$ is the full-array dual-polarization total intensity signal to noise of the calibrator within the single spectral channel of interest.  3C\,84 is known to approach 1\% true polarization at 43\,GHz (e.g.\ \url{http://www.aoc.nrao.edu/~smyers/calibration/master.shtml}); J1407+2827 has been shown to have polarization of $\lessapprox 0.3$\% at Q-band \citep{2018A&A...617A...3L} and $<0.2$\% at K-band \citep{2008A&A...479..409O}.  We adopt $p_{\mathrm{true}}$ = 1\% for 3C\,84 and 0.3\% for J1407+2827; in the case of the 2018 -- 2019 solutions we conservatively adopt $p_{\mathrm{true}}$ = 1\% as the maximum residual with respect to the first-pass fit.  This gives per-channel $p_{\epsilon}$ values $<0.2$\% for 3C\,84, between 0.5 and 3\% for J1407+2827, and between 0.2 and 0.3\% for the combined science targets, where the higher values are at the upper end of the frequency band where the signal-to-noise ratio of most sources decreases.

Also following \citet{2017AJ....154...54H}, we estimate the error in the polarization angle estimates, $\Phi_{\epsilon}$ using their publicly available software \textsc{polcalsims}\footnote{\url{https://github.com/chrishales/polcalsims}} \citep{polcalsims} which uses Monte Carlo simulations to predict position angle error as a function of signal-to-noise of the polarized calibrator.  The per-channel estimates range from 0.15$^{\circ}$ to 1.8$^{\circ}$ in the case of 3C\,48 and 0.05$^{\circ}$ to 0.8$^{\circ}$ in the case of 3C\,286, again with the higher errors corresponding to higher frequencies where the signal-to-noise of the sources is lower.

We assume both of these systematic errors are completely correlated between channels in a spectral window and so use the average per-channel estimate in each spectral window as the overall systematic error for the spectral window estimate.  We use error propagation to translate these systematic errors in $p$ and $\Phi$ in errors in $q$ and $u$:

\begin{align}
\upDelta q_{\mathrm{sys}}^2 &= p_{\epsilon}^2 \cos^2(2\Phi) + 4u^2\Phi_{\epsilon}^2 \nonumber \\
\upDelta u_{\mathrm{sys}}^2 &= p_{\epsilon}^2 \sin^2(2\Phi) + 4q^2\Phi_{\epsilon}^2.
\end{align}

\subsection{$qu$-fitting}\label{S:qu_fitting}

To fit the polarization properties, we take a vector average over time and baseline of the calibrated \emph{RL} and \emph{LR} cross-hand visibility data in each spectral window and solve for $\mathcal{Q}$ and $\mathcal{U}$ using

\begin{align}
V_{RL}&=\mathcal{Q}+i\mathcal{U} \nonumber \\
V_{LR}&=\mathcal{Q}-i\mathcal{U}.
\end{align}\label{eq:crosshands}

We estimate the uncertainty in $q$ and $u$ using the standard deviation in the data and adding in quadrature the systematic errors estimated as in Section~\ref{S:calibration_accuracy} (the error in $\mathcal{I}$ is negligible compared to the error in $\mathcal{Q}$ and $\mathcal{U}$).

We follow a similar $qu$-fitting procedure to \citet{2012MNRAS.421.3300O} in that we fit jointly to $q$ and $u$, minimizing the total $\chi^2$ (rather than fitting to $p$ and $\Phi$ separately).  This guarantees consistency between the derived $p$ and $\Phi$ properties.  Our frequency band does not contain enough information to usefully constrain physical parameters, so unlike \citet{2012MNRAS.421.3300O}, rather than fitting a physical model we simply fit a power-law to the polarization fraction as a function of frequency, and a rotation measure (RM) law, $\Phi = \Phi_{0} + \mathrm{RM} \lambda^{2}$ law to the polarization angle as a function of wavelength.  This gives:

\begin{align}
q = p \cos(2\Phi) &= p_0 \left ( \frac{\nu}{\nu_0} \right ) ^{\alpha_{P}} \cos \left [2 \left (\Phi_0 + \mathrm{RM} \lambda^2 \right ) \right ] \nonumber \\
u = p \sin(2\Phi) &= p_0 \left ( \frac{\nu}{\nu_0} \right ) ^{\alpha_{P}} \sin \left [2 \left (\Phi_0 + \mathrm{RM} \lambda^2 \right ) \right ]
\end{align}\label{eq:qu_def}
where our reference frequency $\nu_0$ is again 34\,GHz.  We use the \textsc{scipy.optimize.leastsq} least-squares fitting algorithm to find the best fitting $p_0$, $\alpha_{P}$, $\Phi_0$ and $\mathrm{RM}$ and estimate their associated errors ($\upDelta p_0$, $\upDelta \alpha_{P}$, $\upDelta \Phi_0$ and $\upDelta \mathrm{RM}$).

We have added the systematic errors to the per-spectral window $q$ and $u$ error estimates before performing the fits to account for small differences in systematic error levels across the frequency band.  However, in general we expect the errors to be correlated across the frequency band and therefore the parameter error estimates (produced assuming uncorrelated errors) will be underestimated.  To account for this, we rescale the $q$ and $u$ errors so that reduced $\chi^2 = 1$ for the fit and re-estimate the parameter errors; then we add in quadrature the mean $p_{\epsilon}$ (to $\upDelta p_0$) and $\Phi_{\epsilon}$ (to $\upDelta \Phi_0$) to produce the final error.

\begin{figure*}
  \begin{center}
     \includegraphics[width=0.33\textwidth, keepaspectratio]{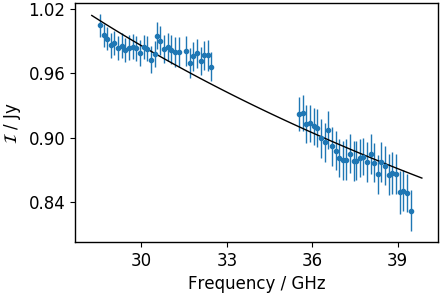}\includegraphics[width=0.33\textwidth, keepaspectratio]{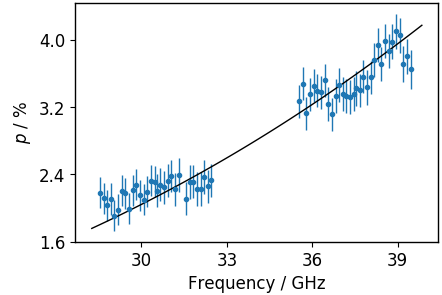}\includegraphics[width=0.33\textwidth, keepaspectratio]{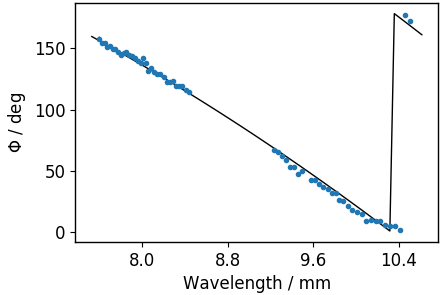}
\caption{Sample fits to the $uv$-plane flux densities: left shows total intensity, centre polarization fraction, and right polarization angle as a function of frequency.  The source is PCCS1\,G147.84-44.04, observed on 10/03/2015, which shows the most extreme rotation measure in the sample.  In all cases, each point represents a spectral window.  In the total intensity plot, the errorbars are derived from the scatter within the frequency channels in the spectral window.  In the polarization plots, the errorbars represent the scatter within the frequency channels added in quadrature with the systematic errors estimated as described in Section~\ref{S:calibration_accuracy}.  The errorbars do not take into account correlations between frequency channels and/or baselines and are not independent.}
\label{Fi:sample_pol_fits}
  \end{center}
\end{figure*}

\begin{figure*}
  \begin{center}
     \includegraphics[width=0.33\textwidth, keepaspectratio]{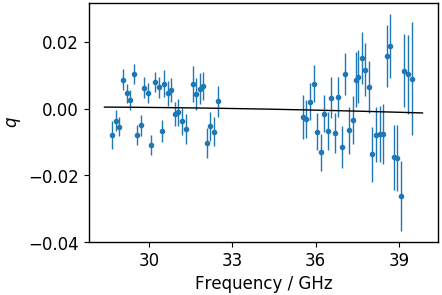}\includegraphics[width=0.33\textwidth, keepaspectratio]{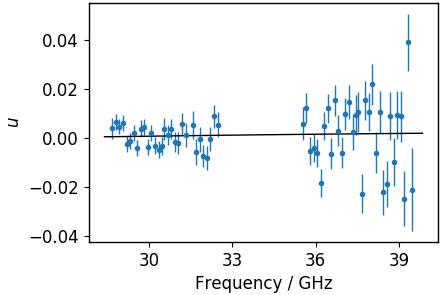}\includegraphics[width=0.33\textwidth, keepaspectratio]{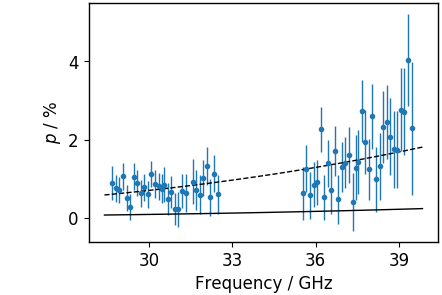}
\caption{Sample fits to the polarization spectrum of J1407+2827, a low-SNR source with negligible polarization.  From left to right we show $q$, $u$ and $p$.  As in Fig.~\ref{Fi:sample_pol_fits}, the errorbars are the total error including statistical and systematic contributions.  The solid black lines show fits to the datapoints using the $qu$-fitting method, while the dashed black line in the $p$ plot shows the bias that would result from fitting directly to $p = \sqrt{q^2 + u^2}$.}
\label{Fi:J1407_polfit}
  \end{center}
\end{figure*}

The $qu$-fitting procedure using our simple model provides reasonable fits in all cases and we do not see systematic trends indicating a more sophisticated model would be warranted; some sample fits are shown in Fig.~\ref{Fi:sample_pol_fits}.  We define as undetected in polarization those sources with fitted $p_0/\upDelta p_0 < 3$.  We note that there are no $n\pi$ ambiguities in the fitted RMs, since our frequency channels are closely spaced.  We note also that this method circumvents the issue of bias in polarization fraction (i.e.\ $p_{\mathrm{meas}} = \sqrt{(q + \sigma_q)^2 + (u + \sigma_u)^2} > p_{\mathrm{true}} = \sqrt{q^2 + u^2}$) since we are fitting a spectrum in $q$ and $u$ across a large number of frequency measurements, each of which can be displaced in either the positive or negative direction.  This is demonstrated in Fig.~\ref{Fi:J1407_polfit} where we show the fit to J1407+2807, the low-SNR source with negligible polarization.  The $qu$-fitting method clearly produces an unbiased result whereas fitting directly to $p = \sqrt{q^2 + u^2}$ produces a biased result.

\section{Results}\label{S:results}

In Table~\ref{T:results} we show a sample section of our results table, containing flux density and spectral index estimates in total intensity and polarization fraction, as well as polarization angles extrapolated to $\lambda = 0$ and RM estimates.

\begin{table*}
\begin{center}
\caption{A sample section of the results table.  Each source has a separate entry for each epoch showing flux density at the central frequency of 34\,GHz and spectral index in total intensity; polarization fraction (not percent) at 34\,GHz and spectral index, extrapolated polarization position angle and RM, along with associated errors.  The errors in the table include the systematic errors estimated as in Section~\ref{S:calibration_accuracy}.  The full table is available as an online-only supplement and at CDS.}
\label{T:results}
\begin{tabular}{lccccccccccccc}
\hline
PCCS1\,030 & Epoch & $I_{0}$ & $\upDelta I_{0}$ & $\alpha$ & $\upDelta \alpha$ & $p_0$ & $\upDelta p_0$ & $\alpha_{P}$ & $\upDelta \alpha_{P}$ & $\Phi_{0}$ & $\upDelta \Phi_{0}$ & RM & $\upDelta$ RM \\ 
 & MJD & Jy & Jy & & & & & & & deg & deg & rad m$^{-2}$ & rad m$^{-2}$ \\ \hline
G178.26+33.40 & 57091.06 & 0.6300 & 0.0315 & -0.360 & 0.101 &  0.02972 & 0.00186 & 0.080 & 0.073 & 108.32 & 1.18 & 391 & 223\\
G182.17+34.17 & 57091.06 & 0.7121 & 0.0356 & -0.426 & 0.101 &  0.06189 & 0.00191 & 0.323 & 0.079 & 44.50 & 0.68 & -789 & 93\\
G200.04+31.88 & 57091.06 & 0.7326 & 0.0366 & 0.058 & 0.101 &  0.01624 & 0.00195 & -0.358 & 0.360 & 31.72 & 1.70 & 3003 & 330\\
G143.53+34.42 & 57091.06 & 2.8432 & 0.1422 & 0.171 & 0.101 &  0.01885 & 0.00133 & -1.187 & 0.139 & 92.04 & 0.86 & 961 & 162\\
G206.82+35.81 & 57091.06 & 5.5528 & 0.2777 & -0.156 & 0.101 &  0.02064 & 0.00186 & 0.551 & 0.101 & 144.15 & 0.93 & 160 & 169\\
G175.72+44.81 & 57091.06 & 2.3493 & 0.1175 & 0.022 & 0.100 &  0.01957 & 0.00186 & -0.793 & 0.105 & 76.35 & 1.53 & -1007 & 301\\
G152.23+41.00 & 57091.06 & 0.8449 & 0.0423 & -0.245 & 0.101 &  0.04056 & 0.00192 & 0.226 & 0.122 & 137.42 & 0.87 & -1529 & 150\\
G198.82+44.43 & 57091.06 & 0.9126 & 0.0456 & 0.238 & 0.100 & <0.00558 & 0.00186 &   - &   - &  - &  - &  - &  -\\
G183.71+46.17 & 57091.06 & 7.3728 & 0.3687 & -0.301 & 0.101 &  0.01772 & 0.00204 & 0.096 & 0.460 & 137.41 & 1.35 & -195 & 240\\
G181.02+50.29 & 57091.06 & 0.5709 & 0.0286 & -0.455 & 0.100 &  0.01440 & 0.00185 & -0.899 & 0.114 & 176.76 & 1.53 & 4664 & 300\\
\hline
\end{tabular}
\end{center}
\end{table*}

\subsection{Total intensity properties}\label{S:intensity_results}

\begin{figure}
  \begin{center}
     \includegraphics[width=0.5\linewidth, keepaspectratio]{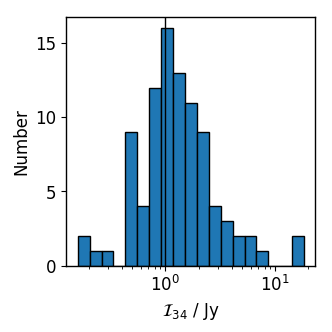}\includegraphics[width=0.5\linewidth, keepaspectratio]{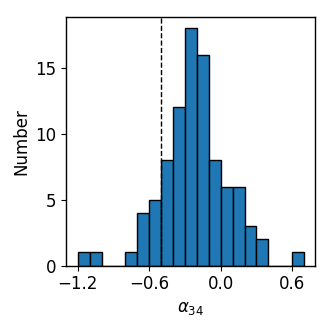}
\caption{Distributions of total intensity flux densities at 34\,GHz (left) and spectral indices (right) across the VLA observation band.  The solid line in the flux density histogram indicates the selection cut at 1\,Jy (at 30\,GHz, but we do not correct for the small frequency shift) and the dashed line in the spectral index histogram indicates the traditional division into steep- and flat-spectrum sources.}
\label{Fi:intensity_hists}
  \end{center}
\end{figure}

In Fig.~\ref{Fi:intensity_hists} we show the distributions of central flux densities (at 34\,GHz) and spectral indices, including each epoch of measurement for each source separately.  These show that most (87\%) of the sources are flat-spectrum with fitted spectral indices $\alpha > -0.5$.  This is as expected given the high-frequency selection criterion and in agreement with the BZCAT classifications.  We also see that although the sources were selected to be brighter than 1\,Jy at 30\,GHz, in fact only slightly more than half (61\%) have 34\,GHz flux densities greater than 1\,Jy; this shows the variability of these sources and highlights the difficulty in selecting a complete sample.

\subsubsection{Total intensity variability}\label{S:intensity_variability}
\begin{figure}
  \begin{center}
     \includegraphics[width=0.5\linewidth, keepaspectratio]{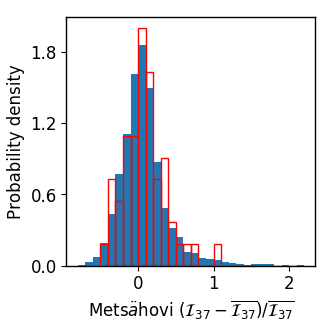}
\caption{Stacked distributions of fractional variations in $\mathcal{I}$ with respect to the mean for sources with Mets\"{a}hovi lightcurves.  The blue, filled histogram shows the total lightcurve while the red, unfilled histogram shows the distribution when sampling each lightcurve only at two points, mimicking the VLA sampling.}
\label{Fi:Mets_variability}
  \end{center}
\end{figure}

Since we only have two epochs of observation for each source, we first use the Mets\"{a}hovi lightcurves to check whether our observations can fairly test the variability of our sample.  For each of the 31 (out of 54) sources with Mets\"{a}hovi measurements spanning $\approx$\,4\,years, we produce a histogram of the `true' variability by stacking all the measurements of $(\mathcal{I} - \bar{\mathcal{I}})/\bar{\mathcal{I}}$ where $\mathcal{I}$ are the individual flux measurements and $\bar{\mathcal{I}}$ is the mean flux density for each source over the whole lightcurve.  Then, we replicate our VLA sampling by interpolating the light curves at points corresponding to the VLA measurements for each source, which produces 55 individual measurements.  We calculate the fractional variation for these points as above, with respect to the mean over the whole lightcurve.  Fig.~\ref{Fi:Mets_variability} shows the resulting two histograms.  They are of course not identical but show very similar distributions, and the standard deviation of each set of measurements is nearly identical, $\sigma=0.30$ and $0.29$.  Restricting the sample to sources for which we have two VLA measurements (24 sources) does not change the distribution or the standard deviation.  We therefore can be confident that, although we only have two epochs of measurement, our sample of sources is large enough to ensure that we are probing the variability statistics of the overall sample fairly.  We note that there is a small asymmetry to larger positive values of the fractional variation apparent in both histograms, which can be explained by the fact that most sources spend a large fraction of their time near a `baseline' flux density value, with occasional excursions to very high, flaring states.  This results in the mean over a long period of time being closer to the baseline, and the flares appearing as large, positive fractional deviations.

\begin{figure}
  \begin{center}
     \includegraphics[width=\linewidth]{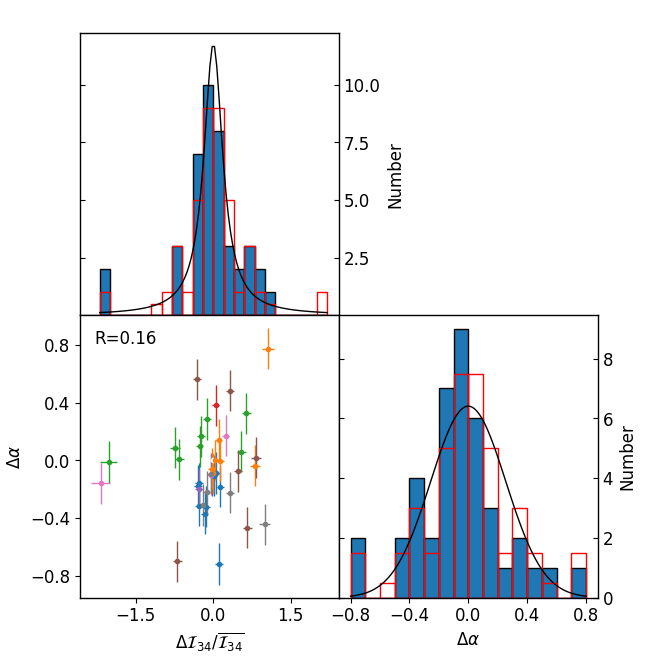}
\caption{Histograms of fractional variation in total intensity flux density and spectral index variation (histograms), and correlation between these quantities colour-coded by epoch pair.  Along with the histograms of the data (solid blue), we also plot the `symmetrized' histograms (red outlines; normalized by a factor of two for comparison with the non-symmetrized histograms) as described in the text.  The black lines show Cauchy (Gaussian) fits to the symmetrized flux density (spectral index) variation distributions; see text for more details.}
\label{Fi:I_variability}
  \end{center}
\end{figure}

Now considering the VLA measurements, in Fig.~\ref{Fi:I_variability} we show histograms of the variability in total intensity flux density and spectral index for the 41 sources for which we have two measurements.  Since we only have two measurements of each source, we define fractional variation in flux density as $(\mathcal{I}_{34,2} - \mathcal{I}_{34,1}) / \overline{\mathcal{I}_{34}}$ where $\overline{\mathcal{I}_{34}}$ is the weighted mean of the two flux densities, and spectral index variation as $\alpha_{2} - \alpha_{1}$.  We attempt to quantify the variability as follows.  We have no reason to expect the flux and spectral index to change more in one direction than the other, given that the sources were observed at different epochs and all will have different variability timescales; we do not expect to see the bias toward positive fractional deviations in $\mathcal{I}$ found in the Mets\"{a}hovi measurements above since our fractional deviations are with respect to the weighted mean of two measurements only, rather than the whole lightcurve.  Therefore we symmetrize the distributions by adding the reflection of each point to the dataset, i.e.\ our symmetrized datasets consist of $\left[ (\mathcal{I}_{34,2} - \mathcal{I}_{34,1}) / \overline{\mathcal{I}_{34}}, (\mathcal{I}_{34,1} - \mathcal{I}_{34,2}) / \overline{\mathcal{I}_{34}} \right]$ and $\left[ \alpha_{2} - \alpha_{1}, \alpha_{1} - \alpha_{2} \right]$.  

To fit a model probability distribution to each of the variability distributions, we use a Kolmogorov--Smirnov (KS) test.  For a given model for the distribution (either Cauchy or Gaussian) and appropriate set of parameters for the model, we calculate the KS `D'-statistic between the symmetrized data and the model which gives a measure of the maximum difference between the cumulative probability distributions of the two.  We then vary the parameters in the model until the smallest `D'-statistic is reached, i.e.\ finding the set of parameters that best match that model to the data.  This method has the advantage that the fit to the model does not depend on how the data are binned.

In the case of the flux density variations, we find a Cauchy distribution centred at 0 fits the distribution well, i.e.\ 
\begin{equation}\label{eq:cauchy}
\mathrm{PDF}\left(\frac{\upDelta {I}_{34}}{\overline{\mathcal{I}_{34}}}\right) = \left \{ \pi \gamma \left [ 1 + \left( \frac{\upDelta {I}_{34}}{\overline{\mathcal{I}_{34}}\gamma} \right)^{2} \right ] \right \}^{-1}
\end{equation}
with the best-fit $\gamma = 0.22$.  The KS test `D'-statistic between the symmetrized (non-symmetrized) data and the fitted Cauchy distribution is 0.04 (0.08) and the $p$-value is 1.00 (0.96) indicating a good fit even to the non-symmetrized data.  We show both the symmetrized and non-symmetrized data in Fig.~\ref{Fi:I_variability}, along with the best-fit Cauchy distribution.

In the case of the $\alpha$ variation, we find a Gaussian distribution to be a better fit than a Cauchy distribution, although clearly it does not fit the data particularly well.   We find a best-fit standard deviation $\sigma=0.26$ with mean $\mu$ fixed to $0$.  The KS test `D' values between the symmetrized (non-symmetrized) data and the fitted distribution are 0.04 (0.18) and the $p$-values are 1.00 (0.13).  The measurements are clearly limited by the relatively large spectral index calibration error, and more accurate measurements with a longer frequency lever arm  would be required to investigate further the spectral index variability.

Fig.~\ref{Fi:I_variability} also shows the correlation between variability in (non-symmetrized) total intensity flux density and spectral index, which is weak with Pearson $R=0.16$, $p$-value=0.33, indicating that there is not enough information in our two-point estimates to investigate the physical mechanisms underlying the changes.  In general one would expect a change in flux density to be accompanied by a change in spectral index and vice versa.  There are, however, some cases where a highly significant variation in spectral index occurs although the flux density is nearly constant; we investigate one of these cases in Section~\ref{S:PCCS9_comp}.   A change in spectral index is therefore also a good indicator of variability.  In total 59\% of the sources have a change in either flux density or spectral index of $>3\sigma$ (using the systematic errors as calibrated in Section~\ref{S:calibration_accuracy}).

\subsection{Polarization properties}\label{S:pol_correlations}

\begin{figure*}
\begin{center}
\includegraphics[width=0.8\textwidth, keepaspectratio]{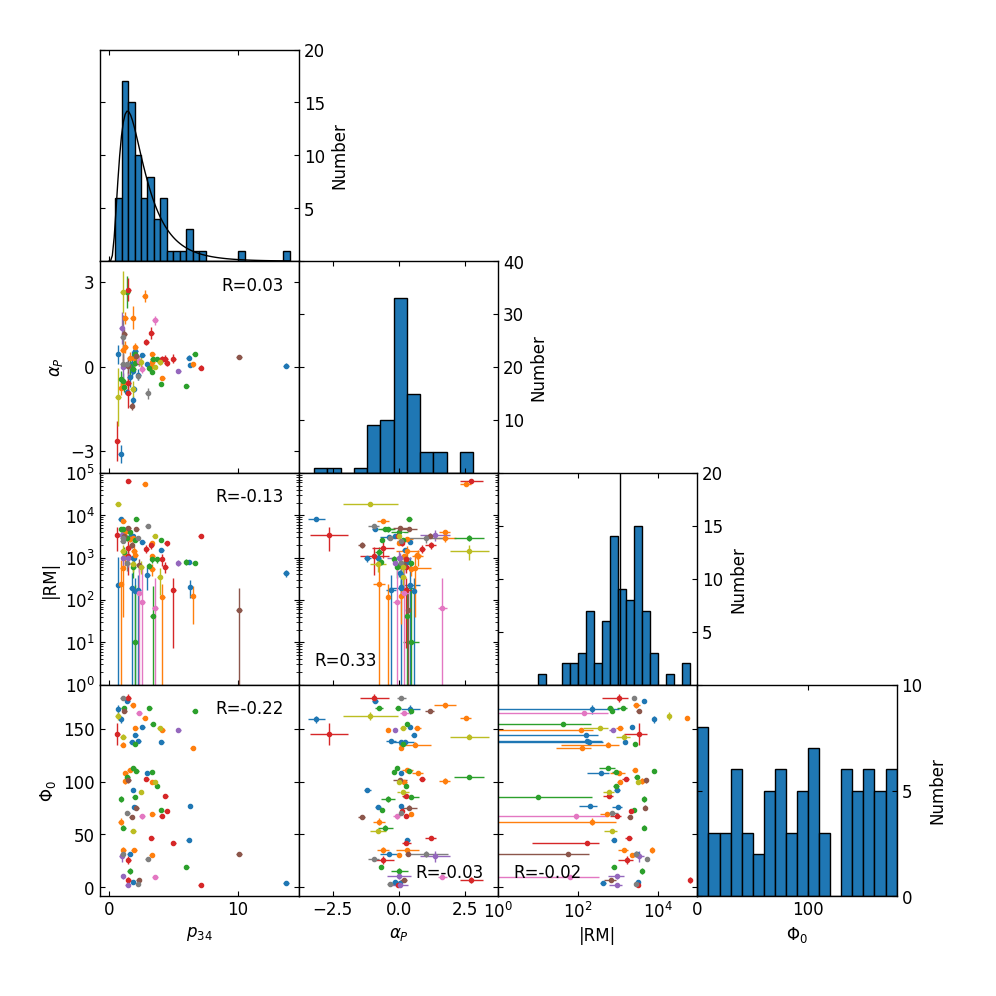}
\caption{Histograms showing the distribution of polarization fractions (in percent), polarization spectral indices, RM (in rad\,m$^{-2}$) and intrinsic polarization angles (in degrees) fit across the VLA bands, and correlations between the four quantities.  The histogram of polarization fractions also shows a log-normal fit to the distribution and the median RM is indicated with a vertical line on the histogram.  Points are colour-coded by observing epoch as in Fig.~\ref{Fi:Mets_OVRO}.}\label{Fi:pol_corr_hists}
\end{center}
\end{figure*}

We detect 85\% of the sources in polarization, considering all observing epochs together.  In Fig.~\ref{Fi:pol_corr_hists} we show the distributions of polarization fractions, spectral indices, RM and intrinsic polarization angles fit to the polarization fraction spectra as well as correlations between the four quantities.  We see fairly low polarizations in most cases, with a median of 2.2\%; this is consistent with other studies at similar frequencies (e.g.\ \citealt{2011ApJ...732...45S}, \citealt{2011MNRAS.413..132B}) and indicates that we are seeing mostly the flat-spectrum core rather than the steep-spectrum lobes, which are more highly polarized.  We also show a log-normal fit to the distribution, made by minimizing the KS `D'-statistic between the data and a log-normal distribution.  This gives fitted parameters $(\mu, \sigma) = (2.20, 0.64)$ which are in good agreement with the fits to flat-spectrum source data at various frequencies from \citet{2018ApJ...858...85P} (e.g.\ their Fig.~3).  We see a fairly broad distribution of spectral indices in polarization (although these do have large errors due to the small change observed over the band), with $\approx$\,26\% of sources having rising spectra at $>3\sigma$ significance indicating depolarization, and $\approx$\,15\% having falling spectra at $>3\sigma$ significance indicating repolarization (we note that repolarization can occur as a consequence of physical models more complicated than a single Faraday screen; see e.g.\ \citet{2012MNRAS.421.3300O} for example models and data).  We do not attempt fits to the $\alpha_P$ and $|\mathrm{RM}|$ distributions due to the large and non-uniform errors.  As expected, the intrinsic polarization angles are uniformly distributed.

We note that the rotation measures in Fig.~\ref{Fi:pol_corr_hists} are observed RMs with no correction for Galactic RM or conversion to the AGN rest frame since we are interested in the observed properties in the context of the QUIJOTE analysis.  We detect an RM at $>3\sigma$ significance in 65\% of the polarization detections and find a median $|\mathrm{RM}| \approx 1100$\,rad\,m$^{-2}$.  Previous VLBI studies have found similarly high AGN core RMs at centimetre wavelengths (e.g.\ \citet{2004ApJ...612..749Z} find core RMs up to $\approx$\,2000 with mean of 644\,rad\,m$^{-2}$ between 8 and 15\,GHz; \citet{2012AJ....144..105H} find core RMs up to $\approx$\,1500 with a tail out to $>6000$ and median of 171\,rad\,m$^{-2}$ between 8 and 15\,GHz) while the jet RMs tend to be lower.  This lends additional support to the idea that our 30 -- 40\,GHz observations are mostly probing the core.  \citet{2019A&A...623A.111H} find that RM increases as a function of frequency in 3C\,273 with RM $\propto \nu^{2}$ in agreement with models for a sheath surrounding a conically expanding flow; if this effect occurs for the majority of sources it would explain our somewhat higher median.  Indeed \citet{2011MNRAS.411...85A} find a median RM of $\approx$\,1600\,rad\,m$^{-2}$ between 15 and 43\,GHz, although their values may suffer from $n\pi$ ambiguities.  We see one extreme outlier in RM which we describe in more detail in Section~\ref{S:PCCS7}.

The only significant correlation between the polarization properties is between the $\alpha_P$ and $|RM|$ measurements, with Pearson $R=0.33$, $p$-value$=0.002$.  This is as expected since a higher RM should imply a greater degree of depolarization at lower frequency, and therefore large, positive $\alpha_P$.  However, the sources with significantly negative $\alpha_P$ also seem to show a high RM, suggesting the situation is more complicated than a single Faraday screen and the polarization angle spectra just happen to follow a $\lambda^2$ law over the relatively small range in wavelength.  Given the uncertainties introduced by averaging over our relatively large beam (compared to VLBI studies which resolve the source structure) we do not investigate this further.

\subsubsection{Polarization variability}

\begin{figure*}
\begin{center}
\includegraphics[width=0.8\textwidth, keepaspectratio]{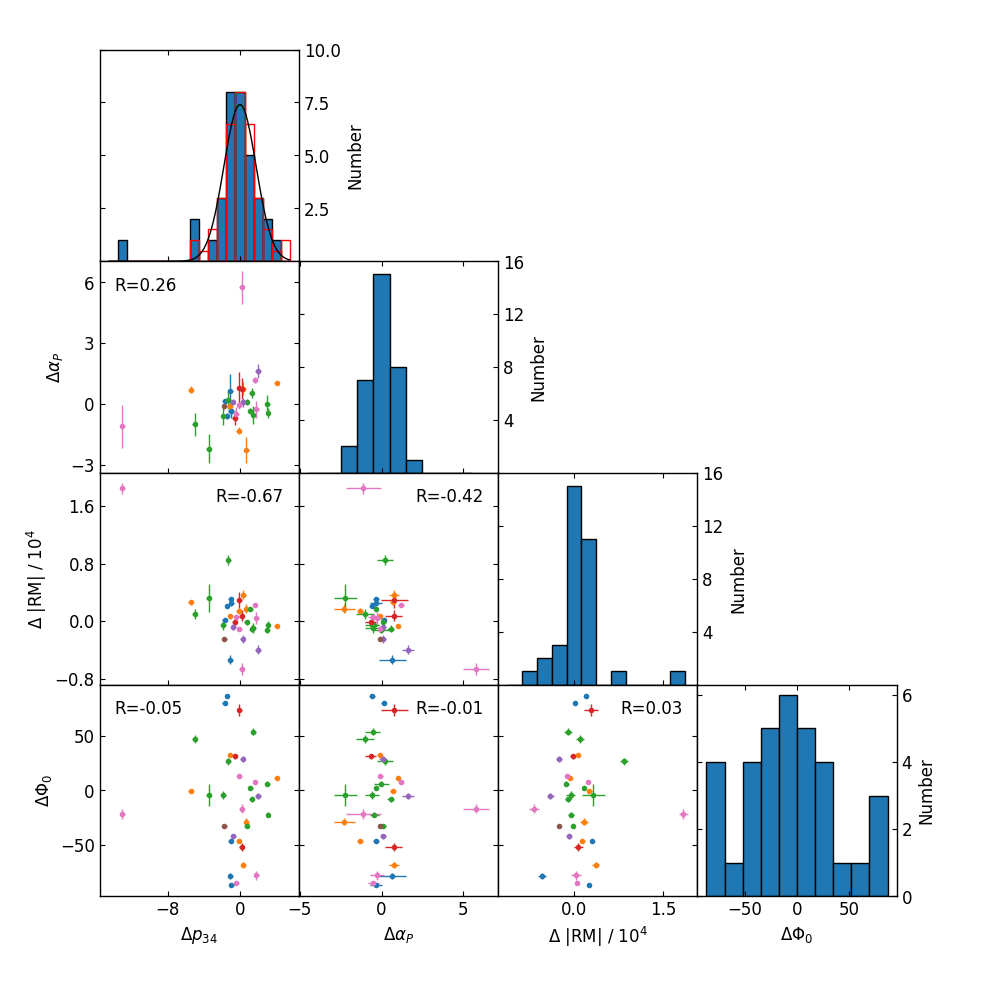}
\caption{Histograms showing the distribution of variation in polarization fraction (in percent), polarization spectral index, RM  (in rad\,m$^{-2}$) and intrinsic polarization angle (in degrees) fit across the VLA bands, and correlations between the four quantities.  Points are colour-coded by observing epoch pair as in Fig.~\ref{Fi:I_variability}.  The histogram of polarization fraction (in percent) also shows a normal distribution centred at 0 with standard deviation equal to the standard deviation measured from the observed distribution, excluding the large outlier.}\label{Fi:P_variability}
\end{center}
\end{figure*}

In Fig.~\ref{Fi:P_variability} we show the distributions of and correlations between the variation in polarization fraction, spectral index, RM and intrinsic polarization angle for the sample.  There are 34 sources with more than one detection; of these 65\% are variable in $p_0$ and all are variable in one or more of these quantities at $3\sigma$ (where $\sigma$ includes systematic and statistical contributions as described in Section~\ref{S:qu_fitting}).  The notable outlier in polarization fraction is PCCS1\,G145.78+43.13 which we investigate further in Section~\ref{S:PCCS34}.  Aside from this extreme outlier, the distribution of changes in polarization fraction is relatively Gaussian; a fit to the symmetrized data excluding the outlier gives a best-fit $\sigma = 1.8$\%.  The KS test comparing the symmetrized (non-symmetrized) data to a normal distribution centred at 0 with this standard deviation gives a $p$-value of 1.00 (0.96) indicating good agreement.

The changes in polarization fraction spectral index and RM are clustered around 0, with the exception of PCCS1\,G145.78+43.13.  It is more difficult to assess the Gaussianity of these distributions given the large variations in measurement errors.  Most of the outliers do have large errors.  Of the two highly significant outliers, one is PCCS1\,G145.78+43.13, as already mentioned and OVRO monitoring of the second, PCCS1\,G129.09-13.46 shows that the two observations similarly took place during a quiescent period and a flare so these large changes are likely genuine.

We see a significant correlation between $\upDelta \alpha_{P}$ and $\upDelta |\mathrm{RM}|$ with Pearson $R=-0.42$ and $p$-value=0.01.  This is in line with the correlation between $\alpha_{P}$ and $|\mathrm{RM}|$ discussed in Section~\ref{S:pol_correlations}; a source undergoing a higher degree of depolarization has a higher RM.  Removing PCCS1\,G145.78+43.13 strengthens the correlation slightly to $R=-0.46$, $p$-value=0.006.  We also see a highly significant correlation between $\upDelta p_{34}$ and $\upDelta |\mathrm{RM}|$ with $R=-0.67$, $p$-value$=1\times 10^{-5}$, however this correlation is largely driven by PCCS1\,G145.78+43.13 and removing this outlier decreases the correlation strength to $R=-0.27$, $p$-value=0.13.  This correlation is also consistent with the idea that a higher RM implies a higher degree of depolarization, so increase in rotation measure correlates with decrease in polarization fraction.

\subsection{Correlations between total intensity and polarization}

\begin{figure}
  \begin{center}
     \includegraphics[width=\linewidth]{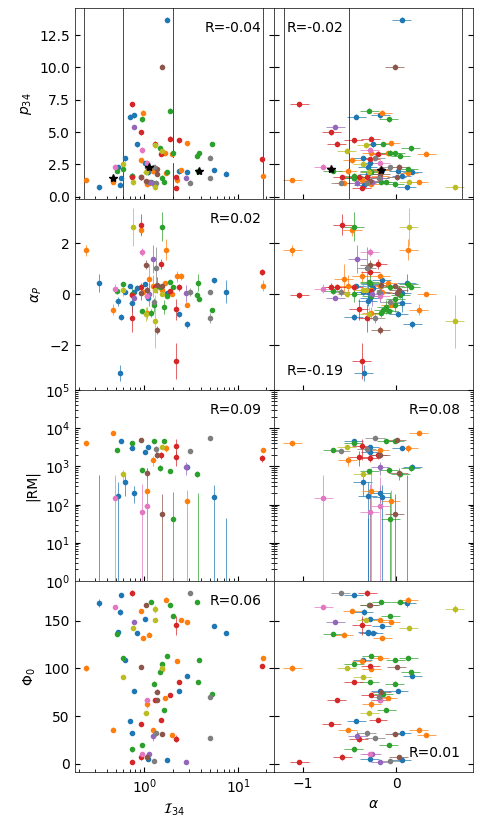}
\caption{Correlations between total intensity flux density (left column) and spectral index (right column) with polarization parameters.  In the $p_{34}$ plots, the vertical lines show a division into flux density/spectral index bins, and the black stars show the median polarization fraction within the bins.}
\label{Fi:I_P_corr}
  \end{center}
\end{figure}

In Fig.~\ref{Fi:I_P_corr} we show the correlation between total intensity parameters and polarization parameters.  All appear uncorrelated.  Of particular importance for predicting the contamination to B-mode analysis is the correlation between total intensity flux density and polarization fraction; i.e.\ if simulating point source contamination, can random values be drawn independently from the total intensity source count and polarization fraction distribution, or is there a correlation?  Although the Pearson $R$ value for $p_{34}$ vs $\mathcal{I}_{34}$ is low ($R = -0.04$, $p  = 0.70$), a correlation could be obscured by the distribution of $\mathcal{I}$ values.  We test this idea by dividing the sources into three bins in $\mathcal{I}$ and calculating median polarization fractions in each bin.  We see a slight positive correlation between flux density and median polarization fraction.  To test the statistical significance of the trend, we calculate KS test statistics between the polarization fractions in each bin and the overall fitted log-normal distribution.  The results for the bins are summarised in Table~\ref{T:I_P_corr}; we find that for the lower and middle flux density bins, the $p$-values are $\approx$\,0.06 and 0.2 respectively, indicating that similar data could be drawn from the overall distribution only $\approx$\,6 and 20\% of the time.  This is contrary to the lack of correlation found by \citet{2011MNRAS.413..132B} and \citet{2008MNRAS.384..775M} although we emphasize that the sample size is small and there could be a selection effect given that our sources were selected to be brighter than 1\,Jy (i.e.\ sources in the lower flux density bin must be very variable and currently in a relatively quiescent state).  The effect is sensitive to the boundary of the lower bin and disappears if, for example, the boundary is placed at 1.0\,Jy rather than 0.6\,Jy.  A sample complete to a lower flux density limit would be required to investigate this further.

We also test the correlation between total intensity spectral index and polarization fraction.  Here we see no evidence for correlation ($R = -0.02$, $p = 0.85$) and no evidence for a different distribution of polarization fractions when we divide into two bins at $\alpha_{34}=-0.5$.  The medians and $p$-values are also reported in Table~\ref{T:I_P_corr}.

\begin{table}
\begin{center}
\caption{Median polarization fraction (in percent) and KS test $p$-values for polarization percentages in different bins in total intensity flux density (top) and spectral index (bottom), compared to the log-normal fit for the overall distribution.  Flux densities are in Jy and $n$ is the number of sources in the bin.  The top row is the overall distribution.}\label{T:I_P_corr}
\begin{tabular}{lcccc}
\hline
$\mathcal{I}_{\mathrm{min}}$ & $\mathcal{I}_{\mathrm{max}}$ & $n$ & $p_{\mathrm{median}}$ & $p$-value \\ \hline
0.23 & 18.4 & 82 & 2.06 & 0.95 \\
0.23 & 0.60 & 9 & 1.44 & 0.060 \\
0.60 & 2.0 & 52 & 2.29 & 0.25 \\
2.0 & 18.4 & 21 & 2.03 & 0.59 \\
\hline
$\alpha_{\mathrm{min}}$ & $\alpha_{\mathrm{max}}$ & $n$ & $p_{\mathrm{median}}$ & $p$-value \\ \hline
-1.2 & 0.7 & 82 & 2.06 & 0.95 \\
-1.2 & -0.5 & 12 & 2.17 & 0.80 \\
-0.5 & 0.7 & 70 & 2.06 & 0.94 \\
\hline
\end{tabular}
\end{center}
\end{table}

\begin{figure}
  \begin{center}
     \includegraphics[width=\linewidth]{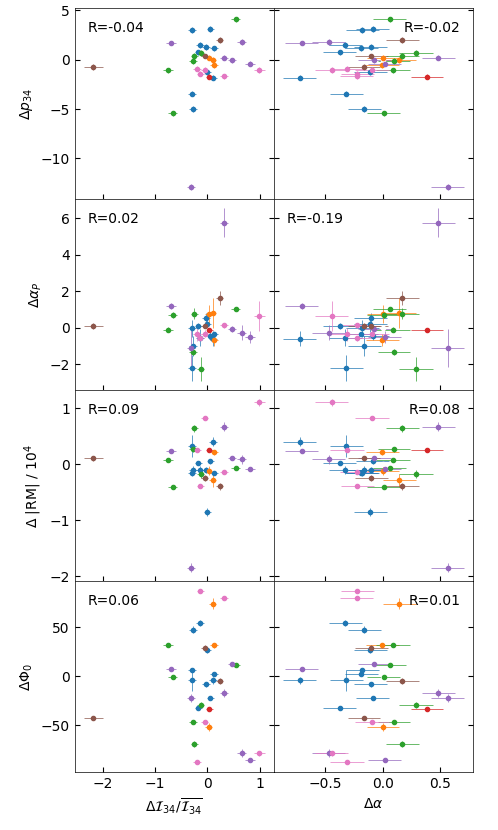}
\caption{Correlations between total intensity flux density variation (left column) and spectral index variation (right column) with variability in the polarization parameters.  Points are coloured by observing epoch pair as in Fig.~\ref{Fi:I_variability}.}
\label{Fi:I_P_var_corr}
  \end{center}
\end{figure}

In Fig.~\ref{Fi:I_P_var_corr} we show correlations between the variations in total intensity flux density and spectral index with variation of the polarization parameters.  Here too we see very little correlation, as evidenced by the very low Pearson $R$-coefficients shown on the plots.  There seems to be a slight correlation between $\upDelta \alpha$ and $\upDelta \alpha_P$ ($R=-0.19$; $p=0.09$) which may indicate that sources coming down from a flare (becoming more optically thin) are also becoming less depolarized ($\alpha_P$ becoming less positive), and vice versa.

\section{Results for some individual sources}\label{S:ind_sources}

Here we investigate in more detail some of the interesting results for individual sources.

\subsection{PCCS1\,G156.86-39.13}\label{S:PCCS9_comp}

PCCS1\,G156.86-39.13 is one of the cases where $\upDelta \alpha$ is large yet its flux density is nearly constant.  It has coverage from OVRO and also at 43\,GHz from the VLBA-BU Blazar Monitoring Project\footnote{\url{http://www.bu.edu/blazars/VLBAproject.html}}.  The VLA data-points are consistently above the VLBA lightcurve since the VLBA resolves out some of the flux from the source, however we can see that the two VLA epochs of observing happen to catch the source on either side of a peak so that the flux density is approximately the same.  Comparing the OVRO and VLBA light-curves we can see that the 43\,GHz flux density decreases more quickly than the 15\,GHz flux density for both this peak and the earlier peak, indicating a change in optical depth.  This is consistent with the VLA spectral indices changing from slightly positive (optically thick) to slightly negative (optically thin) between the two epochs.

\begin{figure}
  \begin{center}
     \includegraphics[width=\linewidth, keepaspectratio]{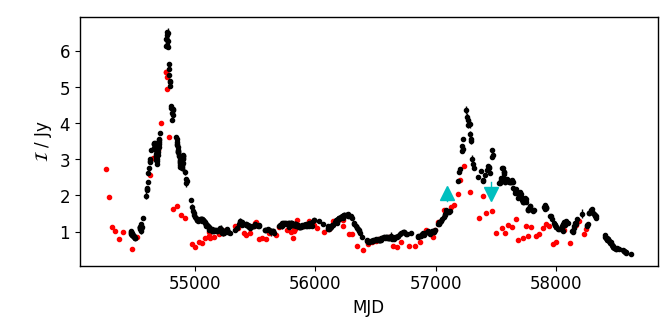}
\caption{Light curves for PCCS1\,G156.86-39.13 from OVRO at 15\,GHz (black errorbars); the VLBA at 43\,GHz (red points) and the VLA observations (cyan lines from 28\,GHz to 40\,GHz, with triangles at the 40\,GHz end).  See text for more detail.}
\label{Fi:PCCS9_comp}
  \end{center}
\end{figure}

\subsection{PCCS1\,G147.84-44.04}\label{S:PCCS7}

PCCS1\,G147.84-44.04 or 4C\,15.05 is an extreme outlier in RM, having RM $\approx -56000$ and $-64000$\,rad\,m$^{-2}$ in the two observation epochs; the polarization fits for the first epoch are shown in Fig.~\ref{Fi:sample_pol_fits}.  At the redshift $z=0.833$ of the source our observed frequency $\nu_{\mathrm{obs}} = 34$\,GHz corresponds to an emitted frequency $\nu_{\mathrm{em}} = (1+z)\nu_{\mathrm{obs}} = 62$\,GHz and the observed RMs correspond to an intrinsic RM of $(1+z)^2 \mathrm{RM}_{\mathrm{obs}} \approx 2 \times 10^5$\,rad\,m$^{-2}$ in the source rest-frame.  To our knowledge this source has not been identified as having a particularly high RM in any other work.  \citet{2004ApJ...612..749Z} and \citet{2012AJ....144..105H} both fail to measure RMs for this source which may be a consequence of the rapid depolarization down to their lower frequencies (see Fig.~\ref{Fi:sample_pol_fits}).  The intrinsic RM is larger than the intrinsic RM measured by \citet{2019A&A...623A.111H} for 3C\,273 of $\approx$\,$3.5\times 10^4$\,rad\,m$^{-2}$ at $\approx$\,62\,GHz (extrapolating the measurement of $6.7\times10^5$\,rad\,m$^{-2}$ at observed wavelength $\approx$\,234\,GHz using RM$_{\mathrm{int}} \propto \nu_{\mathrm{em}}^2$) and $\approx$\,$6\times10^4$\,rad\,m$^{-2}$ found for 3C\,84 (extrapolating the measurement of $8.7\times10^5$\,rad\,m$^{-2}$ at observed wavelength $\approx$\,230\,GHz) from \citet{2014ApJ...797...66P}.  It may even approach the current largest-known RM observed for the lensed quasar PKS 1830-211 \citet{2015Sci...348..311M}, which is $10^8$\,rad\,m$^{-2}$ at $\nu_{\mathrm{em}} = 875$--1050\,GHz, corresponding to $4 \times 10^5$\,rad\,m$^{-2}$ at $\nu_{\mathrm{em}} = 62$\,GHz if the RM$_{\mathrm{int}} \propto \nu_{\mathrm{em}}^2$ law holds over such a wide range in frequency.  

\subsection{PCCS1\,G145.78+43.13}\label{S:PCCS34}
\begin{figure}
  \begin{center}
     \includegraphics[width=\linewidth, keepaspectratio]{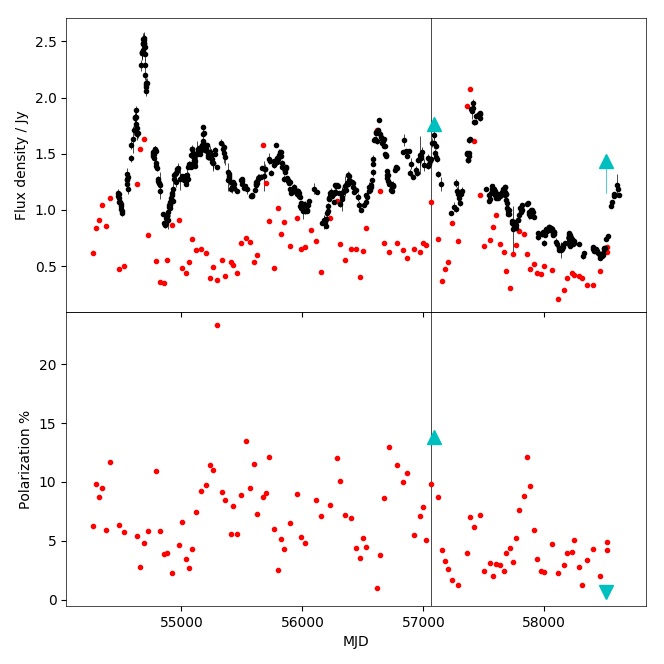}
\caption{Light curves for PCCS1\,G145.78+43.13 from OVRO at 15\,GHz (black errorbars); the VLBA at 43\,GHz (red points) and the VLA observations (cyan lines from 28\,GHz to 40\,GHz, with triangles at the 40\,GHz end).  The top shows total intensity while the bottom shows polarization fraction (in percent).  The vertical black line shows the date of the optical flare.  See text for more detail.}
\label{Fi:PCCS34_lc}
  \end{center}
\end{figure}

PCCS1\,G145.78+43.13 is a notable outlier in polarization fraction variation ($>46\sigma$).  The total intensity and polarization fraction (in percent) light curves for this source are shown in Fig.~\ref{Fi:PCCS34_lc}.  It underwent an exceptionally high optical flare in 2015 (MJD = 57067), coinciding with the emergence of a new knot detected by the VLBA-BU-Blazar Monitoring Project \citep{2018A&A...617A..30M}.  Our first observing epoch of this source happened to coincide with the optical flare, and the polarization fraction measured at this epoch of 13.6\% agrees with the VLBA measurement of $\approx 10$\%.  In the most recent epoch, 28/01/2019, although the total intensity flux density as measured with the VLA has stayed relatively constant at $\approx$\,1.5\,Jy, the total intensity spectral index has changed dramatically from $+0.06$ to $+0.62$; OVRO and VLBA monitoring data confirm that the source is undergoing a radio flare.  The polarization fraction has decreased to 0.7\%, in line with the VLBA polarization values; the polarization fraction spectral index has steepened significantly from $+0.06$ to $-0.92$; and the RM has significantly changed from $-440$ to $-18300$ rad m$^{-2}$. These changes could be attributed to the integrated flux density during the flare state containing a significant contribution from the emerging knot of plasma.

\section{Implications for QUIJOTE}\label{S:implications}

As shown in, e.g.\ \citet{2011MNRAS.413..132B} and \citet{2018ApJ...858...85P}, polarized extra-galactic sources produce significant contamination to the B-mode power spectrum.  The strong variability of the sources that dominate at the QUIJOTE frequencies which will be used for cosmological analysis means that care must be taken to correctly account for their presence.

Following the methodology described in \cite{2012AdAst2012E..52T}, we now estimate the contribution of unresolved polarized radio sources to the angular power spectra at 30\,GHz, and discuss the implications for that frequency channel of the QUIJOTE experiment. For a Poisson distribution of point sources with flux densities below a certain cut-off value $S_{\rm C}$, the contribution to the B-mode angular power spectrum can be estimated as:
\begin{equation}
C_\ell^{BB} = \frac{1}{2} \left( \frac{dB}{dT} \right)^{-2} \langle \Pi^2 \rangle \int_0^{S_{\rm C}} n(S) S^2 dS
\end{equation}
where $n(S)$ is the differential number of sources per steradian, $dB/dT$ is the conversion factor from brightness to temperature, and $\Pi$ corresponds to the fractional polarization.

The model for the differential source counts is taken from \cite{dezotti2005, dezotti2010}\footnote{Available online \url{http://w1.ira.inaf.it/rstools/srccnt/srccnt_tables.html}.}. The average value $\langle \Pi^2 \rangle$ can be computed using the fitted log-normal distribution function in Section~\ref{S:pol_correlations}, using the equations given in \citet{2011MNRAS.413..132B} and \citet{2018ApJ...858...85P}. In our case, we have $\sqrt{\langle \Pi^2 \rangle} \approx 3.3$\,\%. 

\begin{figure}
  \begin{center}
     \includegraphics[width=\linewidth, keepaspectratio]{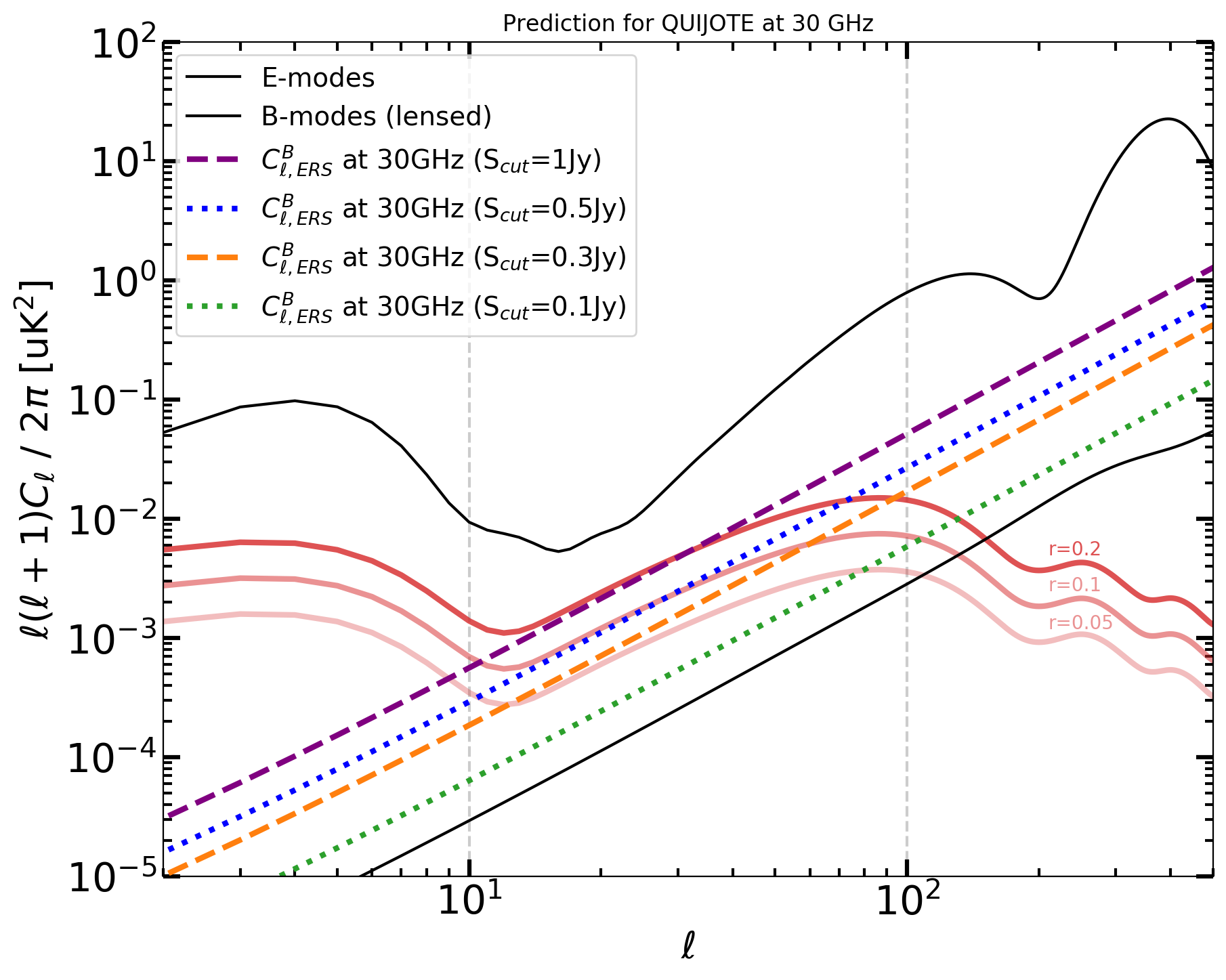}
\caption{Predicted primordial E-mode signal (upper black line), lensing B-mode signal (lower black line), primordial B-mode signal depending on tensor-to-scalar ratio value (red and pink lines, labelled by $r$-value), and point source contamination depending on total intensity cut-off value (coloured dashed and dotted lines, as labelled in caption), all at 30\,GHz. It can be seen that when masking sources with total intensity $>1$\,Jy a similar level of contamination to the $r=0.2$ prediction is reached, while to reach $r=0.05$ sources will likely need to be masked down to 100\,mJy in total intensity.}
\label{Fi:ERS_cont}
  \end{center}
\end{figure}

At 30\,GHz, we obtain $C_\ell^{BB} \approx 1.8\times 10^{-4}$\,$\mu$K$^2$ for $S_{\rm C}=20$\,Jy, and $C_\ell^{BB} \approx 3.2\times 10^{-5}$\,$\mu$K$^2$ for $S_{\rm C}=1$\,Jy. Figure~\ref{Fi:ERS_cont} shows this predicted contribution of radiosources at 30\,GHz, as compared to the expected level of the primordial B-mode signal for values of the tensor-to-scalar ratio of $r=0.2$, $0.1$ and $0.05$. As an indication, we also include the prediction for $S_{\rm C} = 0.3$, $0.5$ and $0.1$\,Jy assuming our polarization fraction results are valid down to these lower flux density limits; we emphasize however that a lower flux density limit would select a different population of sources (not necessarily blazars) and this may not be a valid extrapolation.

If we evaluate the contribution of radiosources to the total power in the BB spectrum at multipole $\ell=80$ (i.e. close to the recombination bump of the expected cosmological signal), we find that the source contribution is reduced from $0.43$\,$\mu$K (for  $S_{\rm C}=20$\,Jy) to $0.18$\,$\mu$K (for $S_{\rm C}=1$\,Jy). This level is comparable to the expected primordial B-mode signal for $r=0.2$, consistently with the results presented in \citet[see their Fig.~7]{2012SPIE.8444E..2YR} and \cite{2005MNRAS.360..935T}.  This highlights the importance of removing the contribution from these sources; since we find that both their total intensity and polarization vary unpredictably, it is likely that they will need to be masked in the analysis rather than subtracted directly, in the absence of simultaneous polarimetric monitoring observations.

Finally, the QUIJOTE experiment aims to reach a limit of $r=0.1$ with measurements at 30\,GHz made with the Thirty-GHz Instrument (TGI), and $r=0.05$ when combining the TGI results with those from the Forty-GHz Instrument \citep{2012SPIE.8444E..2YR}.  As a reference, the best current constraints on tensor-to-scalar ratio are  $r<0.064$ \citep{2020A&A...641A..10P}.  This study suggests that we will need to remove sources down to $\approx$\,500\,mJy in total intensity (see Fig.~\ref{Fi:ERS_cont}) to achieve the $r=0.1$ limit with 30\,GHz data only.  And for the combined analysis of TGI and FGI data, we will possibly need to go below $\approx$\,300\,mJy at 30\,GHz in order to reach the $r=0.05$ goal.  However more studies will need to be done on the source population at these lower flux densities to verify this.  We also emphasize that these results are frequency-dependent and other experiments operating at different frequencies will be differently affected by point source contamination.


\section{Conclusions}\label{S:conclusions}

In order to assess the contamination of the QUIJOTE cosmological fields by polarized emission from radio sources, we observed 51 sources, selected to be brighter than 1\,Jy at 30\,GHz, at 28 -- 40\,GHz with the VLA.  The sample is dominated by flat-spectrum radio quasars.  For 41 of the sources, we have two epochs of observation which allows us to investigate the variability of the sources both in total intensity and polarization.  We find that:
\begin{enumerate}
\item{Our in-band spectral indices agree well with simultaneously measured 15 to 34\,GHz spectral indices using OVRO monitoring data, with some indication of spectral steepening at the higher band.}
\item{The median polarization fraction of our sample at 34\,GHz is 2.2\%, with the largest being 14\%; the distribution of polarization fractions agrees well with polarization fraction distributions at various frequencies summarised in \citet{2018ApJ...858...85P}.}
\item{We find a median rotation measure of |RM|$\approx$\,1110\,rad\,m$^{-2}$, with one extreme outlier (4C\,15.05) having RM $\approx -56000$ and $-64000$\,rad\,m$^{-2}$ in the two observation epochs.  This may be amongst the highest RMs measured up to now in quasar cores.}
\item{We find hints of a correlation between the total intensity flux density and the median polarization fraction, however a larger sample complete to a lower flux density level would be required to confirm this.  We find no correlation between the total intensity spectral index of a source and its polarization fraction.}
\item{59\% of the sources are variable in total intensity, while all are variable in polarization at $3\sigma$ level.  Changes in polarization fraction are roughly Gaussian-distributed with $\sigma=1.8$\%, and one extreme outlier changes by 13\%.}
\item{We find no strong correlations between changes in polarization properties and changes in total intensity flux density or spectral index.}
\item{We conclude that due to these strong variations and lack of correlation, if high-cadence polarimetric monitoring observations of sources at similar frequency are not available, sources must be masked in the QUIJOTE analysis rather than subtracted.  Assuming our results may be extrapolated to lower flux density source populations, sources above $\approx$\,300\,mJy will need to be masked to reach the QUIJOTE goal of $r<0.05$.  For general experiments aiming to detect inflationary B-modes the point source population will need to be studied at the frequency of the experiment to determine the level of masking required.}
\end{enumerate}


\section*{Acknowledgements}

We thank an anonymous reviewer who substantially improved the presentation of the paper.  The {\it QUIJOTE} experiment is being developed by the Instituto de Astrofisica
de Canarias (IAC), the Instituto de Fisica de Cantabria (IFCA), and the
Universities of Cantabria, Manchester and Cambridge. Partial financial support
is provided by the Spanish Ministry of Science, Innovation and Universities
under the projects AYA2007-68058-C03-01, AYA2010-21766-C03-02, AYA2014-60438-P,
AYA2017-84185-P, IACA13-3E-2336, IACA15-BE-3707, EQC2018-004918-P, the Severo
Ochoa Program SEV-2015-0548, and also by the Consolider-Ingenio project
CSD2010-00064 (EPI: Exploring the Physics of Inflation).  This project has received funding from the European Union's Horizon 2020
research and innovation program under grant agreement number 687312
(RADIOFOREGROUNDS).  This publication makes use of data obtained at Mets\"ahovi Radio Observatory, operated by Aalto University in Finland.  This study makes use of 43 GHz VLBA data from the VLBA-BU Blazar Monitoring Program (VLBA-BU-BLAZAR;
http://www.bu.edu/blazars/VLBAproject.html), funded by NASA through the Fermi Guest Investigator Program. The VLBA is an instrument of the National Radio Astronomy Observatory. The National Radio Astronomy Observatory is a facility of the National Science Foundation operated by Associated Universities, Inc. This research has made use of data from the OVRO 40-m monitoring program (Richards, J. L. et al. 2011, ApJS, 194, 29) which is supported in part by NASA grants NNX08AW31G, NNX11A043G, and NNX14AQ89G and NSF grants AST-0808050 and AST-1109911.  YCP is supported by a Trinity College JRF and a Rutherford Discovery Fellowship.  FP acknowledges support from the Spanish Ministerio de Ciencia, Innovaci\'{o}n y Universidades (MICINN) under grant numbers ESP2015--65597-C4-4-R, and ESP2017--86852-C4-2-R. 

\section*{Data availability}

Raw data used in this study are publicly available from the VLA data archive (\url{https://archive.nrao.edu}) and can be accessed using the project codes in Table~\ref{T:obs_dates}.  Calibrated data are available on reasonable request from the authors.


\setlength{\bibsep}{0pt}            
\renewcommand{\bibname}{References} 




\bsp	
\label{lastpage}

\begin{thebibliography}{}

\bibitem[\protect\citeauthoryear{Algaba, Gabuzda, \& Smith}{2012}]{2012MNRAS.420..542A} Algaba J.~C., Gabuzda D.~C., Smith P.~S., 2012, MNRAS, 420, 542 


\bibitem[\protect\citeauthoryear{Algaba, Gabuzda, \& Smith}{2011}]{2011MNRAS.411...85A} Algaba J.~C., Gabuzda D.~C., Smith P.~S., 2011, MNRAS, 411, 85 


\bibitem[\protect\citeauthoryear{Battye et al.}{2011}]{2011MNRAS.413..132B} Battye R.~A., Browne I.~W.~A., Peel M.~W., Jackson N.~J., Dickinson C., 2011, MNRAS, 413, 132 

\bibitem[\protect\citeauthoryear{Bonavera et al.}{2017}]{2017MNRAS.469.2401B} Bonavera L., Gonz{\'a}lez-Nuevo J., Arg{\"u}eso F., Toffolatti L., 2017, MNRAS, 469, 2401 

\bibitem[\protect\citeauthoryear{de Vaucouleurs et al.}{1991}]{1991rc3..book.....D} de Vaucouleurs G., de Vaucouleurs A., Corwin H.~G., Buta R.~J., Paturel G., Fouque P., 1991, rc3..book

\bibitem[\protect\citeauthoryear{de Zotti et al.}{2005}]{dezotti2005} de Zotti G., Ricci R., Mesa D., Silva L., Mazzotta P., Toffolatti L., Gonz{\'a}lez-Nuevo J., 2005, A\&A, 431, 893

\bibitem[\protect\citeauthoryear{de Zotti et al.}{2010}]{dezotti2010} de Zotti G., Massardi M., Negrello M., Wall J., 2010, A\&ARv, 18, 1

\bibitem[\protect\citeauthoryear{Galluzzi \& Massardi}{2016}]{2016IJMPD..2540005G} Galluzzi V., Massardi M., 2016, IJMPD, 25, 1640005 

\bibitem[\protect\citeauthoryear{Galluzzi et al.}{2017}]{2017MNRAS.465.4085G} Galluzzi V., et al., 2017, MNRAS, 465, 4085 

\bibitem[\protect\citeauthoryear{Galluzzi et al.}{2018}]{2018MNRAS.475.1306G} Galluzzi V., et al., 2018, MNRAS, 475, 1306 

\bibitem[\protect\citeauthoryear{Grainge et al.}{2003}]{2003MNRAS.341L..23G} Grainge K., et al., 2003, MNRAS, 341, L23

\bibitem[\protect\citeauthoryear{Hales}{2017a}]{EVLAM201} Hales C.~A., 2017, CASA Interferometric Pipeline Polarization Calibration \& Imaging Requirement \& Design Specifications, EVLA Memo Series, 201; also ALMA Memo Series, 603

\bibitem[\protect\citeauthoryear{Hales}{2017b}]{2017AJ....154...54H} Hales C.~A., 2017, AJ, 154, 54 

\bibitem[\protect\citeauthoryear{Hales}{2017c}]{polcalsims} Hales, C.~A., 2017, Zenodo, http://doi.org/10.5281/zenodo.801336

\bibitem[\protect\citeauthoryear{Hovatta et al.}{2012}]{2012AJ....144..105H} Hovatta T., et al., 2012, AJ, 144, 105

\bibitem[\protect\citeauthoryear{Hovatta et al.}{2019}]{2019A&A...623A.111H} Hovatta T., O'Sullivan S., Mart{\'\i}-Vidal I., Savolainen T., Tchekhovskoy A., 2019, A\&A, 623, A111

\bibitem[\protect\citeauthoryear{Huchra, Vogeley \& Geller}{1999}]{1999ApJS..121..287H} Huchra J.~P., Vogeley M.~S., Geller M.~J., 1999, ApJS, 121, 287

\bibitem[\protect\citeauthoryear{Huffenberger et al.}{2015}]{2015ApJ...806..112H} Huffenberger K.~M., et al., 2015, ApJ, 806, 112 

\bibitem[\protect\citeauthoryear{Jackson et al.}{2010}]{2010MNRAS.401.1388J} Jackson N., Browne I.~W.~A., Battye R.~A., Gabuzda D., Taylor A.~C., 2010, MNRAS, 401, 1388 

\bibitem[\protect\citeauthoryear{Jones et al.}{1985}]{1985ApJ...290..627J} Jones T.~W., Rudnick L., Aller H.~D., Aller M.~F., Hodge P.~E., Fiedler R.~L., 1985, ApJ, 290, 627 

\bibitem[\protect\citeauthoryear{Jorstad et al.}{2007}]{2007AJ....134..799J} Jorstad S.~G., et al., 2007, AJ, 134, 799 

\bibitem[\protect\citeauthoryear{Kravchenko, Cotton, \& Kovalev}{2015}]{2015IAUS..313..128K} Kravchenko E.~V., Cotton W.~D., Kovalev Y.~Y., 2015, IAUS, 313, 128 

\bibitem[\protect\citeauthoryear{Liu et al.}{2018}]{2018A&A...617A...3L} Liu H.~B., Hasegawa Y., Ching T.-C., Lai S.-P., Hirano N., Rao R., 2018, A\&A, 617, A3 

\bibitem[\protect\citeauthoryear{Lister}{2001}]{2001ApJ...562..208L} Lister M.~L., 2001, ApJ, 562, 208 

\bibitem[\protect\citeauthoryear{L{\'o}pez-Caniego et al.}{2009}]{2009ApJ...705..868L} L{\'o}pez-Caniego M., Massardi M., Gonz{\'a}lez-Nuevo J., Lanz L., Herranz D., De Zotti G., Sanz J.~L., Arg{\"u}eso F., 2009, ApJ, 705, 868 

\bibitem[\protect\citeauthoryear{MAGIC Collaboration et al.}{2018}]{2018A&A...617A..30M} MAGIC Collaboration, et al., 2018, A\&A, 617, A30 

\bibitem[\protect\citeauthoryear{Mart{\'\i}-Vidal et al.}{2015}]{2015Sci...348..311M} Mart{\'\i}-Vidal I., Muller S., Vlemmings W., Horellou C., Aalto S., 2015, Sci, 348, 311

\bibitem[\protect\citeauthoryear{Massaro et al.}{2015}]{2015Ap&SS.357...75M} Massaro E., Maselli A., Leto C., Marchegiani P., Perri M., Giommi P., Piranomonte S., 2015, Ap\&SS, 357, 75 

\bibitem[\protect\citeauthoryear{Massardi et al.}{2008}]{2008MNRAS.384..775M} Massardi M., et al., 2008, MNRAS, 384, 775 

\bibitem[\protect\citeauthoryear{Meisner \& Romani}{2010}]{2010ApJ...712...14M} Meisner A.~M., Romani R.~W., 2010, ApJ, 712, 14

\bibitem[\protect\citeauthoryear{Orienti \& Dallacasa}{2008}]{2008A&A...479..409O} Orienti M., Dallacasa D., 2008, A\&A, 479, 409 

\bibitem[\protect\citeauthoryear{O'Sullivan et al.}{2012}]{2012MNRAS.421.3300O} O'Sullivan S.~P., et al., 2012, MNRAS, 421, 3300 

\bibitem[\protect\citeauthoryear{Park et al.}{2018}]{2018ApJ...860..112P} Park J., et al., 2018, ApJ, 860, 112 

\bibitem[\protect\citeauthoryear{Perley \& Butler}{2013}]{2013ApJS..206...16P} Perley R.~A., Butler B.~J., 2013, ApJS, 206, 16 

\bibitem[\protect\citeauthoryear{Perley \& Butler}{2017}]{2017ApJS..230....7P} Perley R.~A., Butler B.~J., 2017, ApJS, 230, 7 

\bibitem[\protect\citeauthoryear{Plambeck et al.}{2014}]{2014ApJ...797...66P} Plambeck R.~L., et al., 2014, ApJ, 797, 66

\bibitem[\protect\citeauthoryear{Planck Collaboration XXVIII}{2014}]{2014A&A...571A..28P} Planck Collaboration XXVIII, 2014, A\&A, 571, A28. 

\bibitem[\protect\citeauthoryear{Planck Collaboration XXVI}{2016}]{2016AA...594A..26P} Planck Collaboration XXVI, 2016, A\&A, 594, A26 

\bibitem[\protect\citeauthoryear{Planck Collaboration XLV}{2016}]{2016A&A...596A.106P} Planck Collaboration XLV, 2016, A\&A, 596, A106 

\bibitem[\protect\citeauthoryear{Planck Collaboration X}{2020}]{2020A&A...641A..10P} Planck Collaboration X, 2020, A\&A, 641, A10. 

\bibitem[\protect\citeauthoryear{Popescu et al.}{1996}]{1996A&AS..116...43P} Popescu C.~C., Hopp U., Hagen H.~J., Elsaesser H., 1996, A\&AS, 116, 43

\bibitem[\protect\citeauthoryear{Puglisi et al.}{2018}]{2018ApJ...858...85P} Puglisi G., et al., 2018, ApJ, 858, 85 

\bibitem[\protect\citeauthoryear{Richards et al.}{2011}]{2011ApJS..194...29R} Richards J.~L., et al., 2011, ApJS, 194, 29 

\bibitem[\protect\citeauthoryear{Rubi{\~n}o-Mart{\'{\i}}n et al.}{2010}]{2010ASSP...14..127R} Rubi{\~n}o-Mart{\'{\i}}n J.~A., et al., 2010, ASSP, 14, 127 

\bibitem[\protect\citeauthoryear{Rubi{\~n}o-Mart{\'\i}n et al.}{2012}]{2012SPIE.8444E..2YR} Rubi{\~n}o-Mart{\'\i}n J.~A., et al., 2012, SPIE,  84442Y, SPIE.8444

\bibitem[\protect\citeauthoryear{Rudnick et al.}{1985}]{1985ApJS...57..693R} Rudnick L., et al., 1985, ApJS, 57, 693 

\bibitem[\protect\citeauthoryear{Sajina et al.}{2011}]{2011ApJ...732...45S} Sajina A., Partridge B., Evans T., Stefl S., Vechik N., Myers S., Dicker S., Korngut P., 2011, ApJ, 732, 45 

\bibitem[\protect\citeauthoryear{Schmidt}{1965}]{1965ApJ...141....1S} Schmidt M., 1965, ApJ, 141, 1

\bibitem[\protect\citeauthoryear{Terasranta et al.}{1992}]{1992A&AS...94..121T} Terasranta H., et al., 1992, A\&AS, 94, 121

\bibitem[\protect\citeauthoryear{Tucci et al.}{2005}]{2005MNRAS.360..935T} Tucci M., Mart{\'\i}nez-Gonz{\'a}lez E., Vielva P., Delabrouille J., 2005, MNRAS, 360, 935

\bibitem[\protect\citeauthoryear{Tucci \& Toffolatti}{2012}]{2012AdAst2012E..52T} Tucci M., Toffolatti L., 2012, AdAst, 2012, 624987

\bibitem[\protect\citeauthoryear{Trager et al.}{2000}]{2000AJ....119.1645T} Trager S.~C., Faber S.~M., Worthey G., Gonz{\'a}lez J.~J., 2000, AJ, 119, 1645

\bibitem[\protect\citeauthoryear{Zavala \& Taylor}{2004}]{2004ApJ...612..749Z} Zavala R.~T., Taylor G.~B., 2004, ApJ, 612, 749

\end{thebibliography}
\end{document}